\documentclass[acmsmall,screen,nonacm,authorversion]{acmart}

\usepackage{cleveref}
\usepackage{subcaption}

\usepackage[frozencache=true,outputdir=build,cachedir=minted-cache]{minted}
\setminted{fontsize=\footnotesize}
\setmintedinline{fontsize=\normalsize}

\usepackage[utf8x]{inputenc}
\usepackage{hyphenat}
\setcitestyle{nosort}
\iftrue % use \iftrue / \iffalse to turn on/off comments
  \newcommand{\newcommenter}[3]{%
    \newcommand{#1}[1]{%
      \textcolor{#2}{\small\textsf{[{#3}: {##1}]}}%
    }%
  }
\else
  \newcommand{\newcommenter}[3]{\newcommand{#1}[1]{}}
\fi

\newcommenter{\lk}{blue}{LK}
\newcommenter{\sk}{teal}{SK}
\newcommenter{\sz}{orange}{SZ}

\newcommand{\hs}[1]{\mintinline{haskell}{#1}}
\newcommand{\rs}[1]{\mintinline{rust}{#1}}

\newcommand{\code}[1]{\mintinline{rust}{#1}}

\newcommand{\at}{\mathbin{@}}
\newcommand{\tr}[1]{\textrm{#1}}
\newcommand{\ts}[1]{\textsf{#1}}
\newcommand{\tsb}[1]{\textbf{\textsf{#1}}}
\newcommand{\ti}[1]{\textit{#1}}

\begin{document}

\title{Portable, Efficient, and Practical Library-Level Choreographic Programming}

\author{Shun Kashiwa}
\orcid{0009-0001-3665-0182}
\affiliation{%
  \institution{University of California, Santa Cruz}
  \country{USA}
  \city{Santa Cruz}}
\email{shkashiw@ucsc.edu}

\author{Gan Shen}
\orcid{0009-0006-0947-9531}
\affiliation{%
  \institution{University of California, Santa Cruz}
  \country{USA}
  \city{Santa Cruz}}
\email{gshen42@ucsc.edu}

\author{Soroush Zare}
\orcid{0009-0003-5558-4111}
\affiliation{%
  \institution{University of California, Santa Cruz}
  \country{USA}
  \city{Santa Cruz}}
\email{sozare@ucsc.edu}

\author{Lindsey Kuper}
\orcid{0000-0002-1374-7715}
\affiliation{%
  \institution{University of California, Santa Cruz}
  \country{USA}
  \city{Santa Cruz}}
\email{lkuper@ucsc.edu}

\renewcommand{\shortauthors}{Kashiwa et al.}

\begin{abstract}
  Choreographic programming (CP) is an emerging paradigm for programming distributed applications that run on multiple nodes.
  In CP, instead of implementing individual programs for each node, the programmer writes one, unified program,
  called a \emph{choreography}, that is then transformed to individual programs for each node via a compilation step called endpoint projection (EPP).
  While CP languages have existed for over a decade, \emph{library-level} CP
  --- in which choreographies are expressed as programs in an existing host language, and choreographic language constructs
  and endpoint projection are provided entirely by a host-language library ---
  is in its infancy.  Library-level CP has great potential, but the only existing implementation approaches have portability, efficiency, and practicality drawbacks that hinder its adoption.

  In this paper, we aim to advance the state of the art of library-level CP with two novel techniques for choreographic library design and implementation: \emph{endpoint projection as dependency injection} (EPP-as-DI), and \emph{choreographic enclaves}.
  EPP-as-DI is a language-agnostic technique for implementing EPP at the library level. Unlike existing library-level approaches, EPP-as-DI
  asks little from the host language --- support for higher-order
  functions is all that is required --- making it usable in a wide
  variety of host languages.
  Choreographic enclaves are a language feature that lets the programmer define \emph{sub-choreographies} within a larger choreography.  Within an enclave, ``knowledge of choice'' is propagated only among the enclave participants, enabling the seamless use of the host language's conditional constructs while addressing the efficiency limitations of existing library-level implementations of choreographic conditionals.
  We implement EPP-as-DI and choreographic enclaves in ChoRus, the first CP library for the Rust programming language.  Our case studies and benchmarks demonstrate that the usability and performance of ChoRus compares favorably to traditional, non-choreographic distributed programming in Rust.
\end{abstract}

\maketitle

\section{Introduction}
\label{sec:intro}

In a distributed system, a collection of independent nodes communicate with each other by sending and receiving messages.
Programmers must ensure that nodes' local behaviors --- sending and receiving messages, and taking internal actions --- together amount to the desired global behavior of the entire system.
As a very simple example, consider a distributed protocol involving nodes Alice and Bob, in which Alice sends a greeting to Bob and Bob responds.
In traditional distributed programming (assuming the existence of \code{send} and \code{receive} functions that implement message transport), Alice might run a node-local program \code{send("Hello!", Bob); receive(Bob)}.
Meanwhile, Bob would run his own node-local program \code{receive(Alice); send("Hi!", Alice)}.
Alice and Bob depend on each other to faithfully follow the protocol: if either of them forgets to call \code{send}, for instance, then their counterpart will wait forever to receive a message (or time out and report an error).
This approach is prone to bugs, including deadlocks.

The emerging paradigm of \emph{choreographic programming}~\citep{deadlock-freedom-by-design, montesi-dissertation, cruz-filipe-montesi-core, giallorenzo-choral, hirsch2022pirouette, HasChor, montesi-book} offers a way to rule out this class of bugs.
Instead of programming individual nodes, the choreographic programmer writes a single program, called a \emph{choreography}, that expresses the behavior of the entire system from an objective, third-party point of view.
For example, the above protocol might be written as the choreography \code{Alice("Hello!") ~> Bob;} \code{Bob("Hi!") ~> Alice}.  The \code{~>} operator denotes communication between a sender and a receiver.
Choreographies are transformed into collections of node-local programs via a compilation step called \emph{endpoint projection} (EPP)~\citep{qiu-foundation,carbone-cdl-epp-esop,carbone-cdl-epp}.
If EPP is correct, every \code{send} in one of the resulting node-local programs is guaranteed to have a corresponding \code{receive} in another node-local program, ensuring deadlock freedom~\cite{deadlock-freedom-by-design}.

In the last ten years, several choreographic programming (CP) languages have been proposed~\citep{deadlock-freedom-by-design,montesi-dissertation,dalla2014aiocj,aiocj,giallorenzo-choral,hirsch2022pirouette}.
However, \emph{library-level} CP --- in which choreographies are expressed as programs in an existing host language, and choreographic operators and EPP are provided entirely by a host-language library --- is just beginning to emerge.

Library-level CP has the potential to immensely improve the accessibility and practicality of CP by meeting programmers where they are --- in their programming language of choice, with access to that language's ecosystem.
A library-level implementation of choreographic programming in a given host language would enjoy the usual advantages of embedded DSLs~\citep{hudak-edsls}: it would be installable just like any host-language library, compilable just like any host-language program, and could use any host-language-specific tools for development, debugging, and deployment.
Library-level CP would also aid the integration of choreographies into larger systems, without any need for the programmer to change languages just to implement certain components choreographically.  Library-level CP would fit especially nicely into a workflow in which a programmer wishes to port a \emph{non}-distributed program --- say, a turn-based game in which players sit next to each other at the same machine --- to a distributed implementation in which the players interact over a network.  With CP, we start and end the process with \emph{one} program, and library-level CP further means that we need never switch languages along the way.

Given these advantages, how can we accomplish library-level CP?
So far, the only existing library-level CP implementation is the recently proposed HasChor framework~\citep{HasChor}, which implements support for CP by means of a domain-specific language embedded in Haskell.
In HasChor, choreographies are monadic computations in which choreographic operators such as \code{~>} may be used.
Under the hood, the HasChor library uses a clever implementation technique based on dynamic interpretation of \emph{freer monads}~\citep{kiselyov-more-ext-effs} to carry out EPP.

The HasChor framework represents the current state of the art of library-level CP.
However, HasChor's implementation approach has several limitations that hinder its portability, efficiency, and practicality.
First, HasChor's implementation relies on Haskell-specific language features and is not easily portable to other languages that lack Haskell's particular arsenal of programming abstractions. In particular, HasChor's use of freer monads to implement EPP makes it challenging to port to other languages.
Second, HasChor's implementation of EPP --- which is the only existing library-level implementation of EPP for choreographies --- results in inefficient runtime behavior in node-local programs compared to what standalone choreographic languages can offer.
Specifically, HasChor's treatment of conditionals in choreographies results in unnecessary network traffic, making it unsuitable for use in typical distributed deployments in which network bandwidth is a scarce resource.  The HasChor approach also requires programmers to use a HasChor-specific language construct for conditionals, rather than using the control flow constructs of the host language.
Finally, HasChor lacks features that would make it easier to integrate choreographic components into larger, non-choreographic software systems.
This is unfortunate, since the ability to seamlessly integrate choreographic and non-choreographic code should be a selling point of the library-level CP approach, as opposed to standalone CP languages.

In this paper, we aim to advance the state of the art of library-level CP by addressing the above limitations.
We make the following specific contributions:
\begin{itemize}
  \item We propose \emph{endpoint projection as dependency injection} (EPP-as-DI), a novel and language-agnostic implementation technique for library-level CP~(\Cref{sec:epp-as-di}).  Unlike the HasChor implementation approach, EPP-as-DI asks little from the host language: support for \emph{higher-order functions} is all that is required.  As such, the EPP-as-DI approach is straightforward to use in a wide variety of host languages.
  \item We propose a novel design and implementation technique for implementing efficient conditionals in library-level CP: \emph{choreographic enclaves}~(\Cref{sec:enclaves}).  Using enclaves, a programmer can sidestep the bandwidth inefficiency of a naive implementation of choreographic conditionals, while still making seamless use of the host language's conditional constructs.
  \item We present ChoRus, a choreographic programming library for Rust, implemented using our proposed techniques~(\Cref{sec:chorus}).  Along with EPP-as-DI and choreographic enclaves, ChoRus is the first CP library to support \emph{located} arguments and return values~(\Cref{subsec:chorus-located-io}), a feature that aids the integration of choreographic components into larger, non-choreographic software projects.  We empirically evaluate the usability and performance of ChoRus compared to traditional distributed programming in Rust~(\Cref{sec:evaluation}).
\end{itemize}
The ChoRus implementation, case studies and benchmarking code, and documentation are available at \url{https://github.com/lsd-ucsc/ChoRus}.

% !TEX root = ./main.tex

\section{Background on Choreographic Programming}
\label{sec:cp-in-a-nutshell}

We begin in \Cref{subsec:choreographic-approach} with a brief overview of CP using an example implemented in the standalone CP language Choral~\citep{giallorenzo-choral}.
Then, in \Cref{subsec:llcp}, we give an overview of library-level CP using the HasChor framework~\cite{HasChor}, and we discuss the strengths and limitations of library-level CP as it stands today.
For a comprehensive introduction to CP, we refer the reader to~\citet{montesi-book,montesi-dissertation}.

\subsection{The Elements of Choreographic Programming}
\label{subsec:choreographic-approach}

To illustrate the key concepts of CP, let us consider a well-known example from the literature: the ``bookseller'' protocol~\citep{carbone-cdl-epp-esop,carbone-cdl-epp,honda-mpsts,w3c-cdl-primer}.
This protocol describes the interactions between a bookseller and a potential book buyer.
First, the buyer sends the name of a book they wish to purchase to the seller.
In response, the seller looks up the catalog and sends back the price of the book to the buyer, who then checks whether the price is within their budget.
If the buyer has the means to purchase the book, they notify the seller and obtain an estimated delivery date.
Alternatively, if the book's cost exceeds their budget, they communicate to the seller their decision not to proceed with the purchase.
\Cref{fig:bookseller-individual} shows how this protocol might be implemented in a traditional (non-choreographic) fashion as two individual programs, running on the buyer and seller's distinct nodes.
We use Python in \Cref{fig:bookseller-individual} as a representative mainstream programming language.
We assume that the \texttt{send} and \texttt{receive} functions are provided by some library that implements network communication between nodes.

\begin{figure}[t]
    \begin{minipage}[t]{0.2\textwidth}
        \begin{minted}[linenos]{python}
def buyer():
    title = input()
    send(title, "seller")
    price = receive("seller")
    decision = price <= budget
    if decision:
        send(True, "seller")
        delivery = receive("seller")
    else:
        send(False, "seller")
        \end{minted}
    \end{minipage}
    \hfill
    \begin{minipage}[t]{0.5\textwidth}
        \begin{minted}[linenos]{python}
def seller():
    title = receive("buyer")
    price = catalog.get_price(title)
    send(price, "buyer")
    decision = receive("buyer")
    if decision:
        delivery = catalog.get_delivery(title)
        send(delivery_date, "buyer")
        \end{minted}
    \end{minipage}
    \caption{The bookseller protocol implemented as individual node-local programs}
    \label{fig:bookseller-individual}
    \Description{The bookseller protocol implemented as individual node-local programs}
\end{figure}

Even for simple protocols like the bookseller protocol, it is easy to introduce bugs.
For example, the programmer might forget to send the decision to the seller, in which case the seller will wait indefinitely for the buyer's response, causing a deadlock.
The programmer might also use different encodings for the delivery date in the buyer and seller programs, in which case the buyer will not be able to parse the delivery date sent by the seller, causing a type error.

Choreographic programming addresses these problems by letting the programmer implement a protocol as a single, unified program, called a choreography.
\Cref{fig:bookseller-choral} shows an implementation of the bookseller protocol as a choreography in Choral~\citep{giallorenzo-choral}, a standalone CP language.
\begin{figure}
    \begin{minted}[linenos]{java}
String@Buyer title_buyer = UI@Buyer.input();
String@Seller title_seller = c.<String>com(title_buyer);
Integer@Buyer price = c.<Integer>com(catalog.quote(title_seller));
boolean@Buyer decision = price <= budget;
if(decision) {
    c.<EnumBoolean>select(EnumBoolean@Buyer.True);
    String@Seller delivery = catalog.get_delivery(title_seller);
    String@Buyer delivery2 = c.<String>com(delivery);
} else {
    c.<EnumBoolean>select(EnumBoolean@Buyer.False);
}
    \end{minted}
    \caption{The bookseller protocol implemented in the Choral choreographic language}
    \Description{The bookseller protocol implemented in the Choral choreographic language}
    \label{fig:bookseller-choral}
\end{figure}
Taking \Cref{fig:bookseller-choral}'s implementation of the bookseller protocol as an example, let us consider four key elements of CP:
\begin{itemize}
    \item \textbf{Located data and computation.} The bookseller protocol involves two locations, \code{Buyer} and \code{Seller}, and data and computation reside at one of these locations. On line 1 of \Cref{fig:bookseller-choral}, the call to the function \code{input} happens at the buyer, as indicated by the \code{@Buyer} annotation on the function call. Likewise, the value returned by \code{input} is located at the buyer, which we see in its type, \code{String@Buyer}. Choral's type system ensures that data at one location cannot be accessed at a different location without an explicit communication.\footnote{While choreographic languages often represent locations at the type level, the notion of located data and computation is always present in choreographic programming, whether or not locations are made explicit in a type system like Choral's.}
    \item \textbf{A unified language construct for communication.} Choreographies replace explicit calls to \code{send} and \code{receive} with a single language construct representing communication between a sender and a receiver.  In the bookseller protocol, the buyer sends the title of the book to the seller (and the seller receives it) on line 2 of \Cref{fig:bookseller-choral}, using the \code{com} method. Here, \code{com} takes a string located at the buyer and returns a string located at the seller.  Additional calls to \code{com} on lines 3 and 8 express communications from the seller to the buyer.
    \item \textbf{Propagation of ``knowledge of choice''.}  On line 4 of \Cref{fig:bookseller-choral}, the buyer checks whether the price of the book is within their budget, and depending on the decision, the choreography takes different branches. Conditionals in choreographic programming are challenging because of the problem known as ``knowledge of choice''~\citep{castagna-knowledge-of-choice}. When the branches of a conditional expression encode different communication patterns, all affected locations must be notified of the outcome of evaluating the conditional. In Choral, the \code{select} method is used to express \emph{selections}, which indicate that the choreography has taken a particular branch and propagate the information to relevant locations. On line 6, the buyer uses \code{select} to send \code{True} to the seller if the price is within the budget; otherwise, the buyer sends \code{False} to the seller on line 10. The seller will receive \code{True} or \code{False} and take the appropriate branch.
    \item \textbf{Endpoint projection.} By itself, a choreography is useful as a global specification of the behavior of a protocol.  If we wish to have a runnable implementation, however, we need a way to perform \emph{endpoint projection} (EPP).\footnote{Unlike in the literature on multiparty session types~\citep{honda-mpsts}, in which endpoint projection refers to projecting a \emph{global type} to a collection of \emph{local types}, in choreographic programming we are concerned with projecting a \emph{global program} (that is, a choreography) to a collection of \emph{local programs}.} EPP  transforms a choreography into an individual program for each target node. For the bookseller protocol, the Choral compiler carries out EPP and generates Java programs similar to those in \Cref{fig:bookseller-individual}.
\end{itemize}

So far, we have been using Choral to illustrate the key concepts of CP.
Choral exemplifies \emph{language-level} CP, where choreographies are programs in a standalone language with its own syntax, type system, compiler, and so on.
Nearly all existing choreographic programming languages are implemented as standalone languages.
\Cref{sec:related-work} discusses Choral and other standalone choreographic languages in more detail.
We now turn to \emph{library-level} CP, the focus of this paper.

\subsection{CP Implemented as a Library}
\label{subsec:llcp}

As discussed in \Cref{sec:intro}, the only existing library-level CP framework is HasChor~\citep{HasChor}, which implements CP as a Haskell library.
\Cref{fig:bookseller-haschor} shows \citeauthor{HasChor}'s implementation of the bookseller protocol as a Haskell program using HasChor.

\begin{figure}[t]
    \centering
    \begin{minted}[xleftmargin=0.6cm, linenos]{haskell}
bookseller :: Choreo IO (Maybe Day @ "buyer")
bookseller = do
  title'   <- (buyer, title) ~> seller
  price    <- seller `locally` \un -> return (priceOf (un title'))
  price'   <- (seller, price) ~> buyer  
  decision <- buyer `locally` \un -> return (un price' <= budget)
  
  cond (buyer, decision) \case
    True -> do
      date  <- seller `locally` \un -> return (deliveryDate (un title'))
      date' <- (seller, date) ~> buyer
      buyer `locally` \un -> return $ Just (un date')
    False -> do
      buyer `locally` \_ -> return Nothing
    \end{minted}
    \caption{The bookseller protocol implemented in Haskell using HasChor~\citep{HasChor}}
    \label{fig:bookseller-haschor}
\end{figure}

With HasChor, choreographies are written as computations that run in the \hs{Choreo} monad provided by the library.  The \hs{bookseller} choreography's type signature on line 1 of \Cref{fig:bookseller-haschor} shows that it returns a value of type \hs{Maybe Day} at the \hs{"buyer"} location.  In general, HasChor supports \emph{located values} of type \hs{a @ l}, implemented using GHC Haskell's support for type-level symbols.
The \hs{(~>)} operator, seen on lines 3, 5, and 11 of \Cref{fig:bookseller-haschor}, implements communication between a sender and a receiver and is HasChor's counterpart of the \code{com} method in Choral.
The \hs{locally} operator, on lines 4, 6, 10, 12, and 14 of \Cref{fig:bookseller-haschor}, implements local computation at a particular node --- for instance, looking up the book's price on the seller's node (line 4), and computing whether the book is in budget on the buyer's node (line 6).  A located value of type \hs{a @ l} may be ``unwrapped'' and used at the specified location \hs{l} using the special \hs{un} function passed to \hs{locally}.
Finally, the \hs{cond} operator on line 8 of \Cref{fig:bookseller-haschor} implements a choreographic conditional expression.
Unlike with the Choral bookseller implementation in \Cref{fig:bookseller-choral}, the HasChor programmer does not need to use anything like \code{select} to solve the knowledge-of-choice problem.
Instead, in HasChor, \hs{cond} \emph{automatically} inserts the necessary communication to propagate knowledge of choice.
While this design choice saves the programmer the tedium of writing calls to \code{select}, it has unfortunate consequences for efficiency, as we will discuss in \Cref{sec:enclaves}.

To run \hs{Choreo} computations, the HasChor framework provides a \hs{runChoreography} function that performs endpoint projection.  Given a choreography of \hs{Choreo} type (such as \hs{bookseller}) and a location name (such as \hs{"buyer"}), \hs{runChoreography} acts something like a just-in-time compiler: it dynamically generates (and runs) a node-local program at the specified location, by dynamically interpreting the choreography.  This approach to library-level EPP is possible because HasChor implements \hs{Choreo} as a \emph{freer monad}~\citep{kiselyov-more-ext-effs}, whose operations can be given different semantics depending on the location at which they are run.  For instance, in HasChor the \code{~>} operator is interpreted as \code{send} for the sender, \code{receive} for the receiver, and as a no-op for other participants in a choreography.

Library-level CP has the usual advantages of embedding a DSL in an existing host language, including ability to piggyback on the host language's ecosystem and tooling, a gentle learning curve for host-language users, and seamless integration with existing host-language code.  HasChor enjoys all of these advantages.
It is therefore tempting to directly port the HasChor library to lots of languages in which programmers might benefit from CP.
A world with PyChor, JSChor, JavaChor and RustChor libraries would surely make CP more practical and accessible than it is today.
Unfortunately, this ``port HasChor to your favorite language'' plan has some flaws:
\begin{itemize}
    \item \emph{Tight coupling with Haskell and monads.}  HasChor's monadic implementation approach relies on Haskell-specific language features.  While these implementation choices are appropriate (and elegant) in the context of Haskell, they are not necessarily easily portable to other languages. To make choreographic programming more widely accessible, a more general approach to implementing library-level CP is called for.
    \item \emph{Inefficient conditionals.} HasChor's implementation of conditionals in choreographies involves broadcasting the value of the condition expression to \emph{all} nodes participating in the choreography, even those nodes that are not involved in the execution of the conditional.  Implementing conditionals efficiently is a particular challenge for library-level CP: while standalone choreographic languages can statically analyze choreographies to insert only the minimum ammount of inter-node communication needed, such an analysis would be difficult (if not impossible) to accomplish in HasChor, given its implementation approach that relies on dynamic interpretation of free monads.  Therefore, HasChor's implementation of conditionals is unlikely to scale well to systems with large numbers of nodes, or those where network bandwidth is a bottleneck.
    \item \emph{Lack of support for located arguments and return values}. One of the biggest advantages of library-level CP is that it can easily be integrated with existing host-language code. However, HasChor does not support providing located arguments to choreographies or returning located values from choreographies. This limitation makes it difficult to use HasChor as part of a larger application. 
\end{itemize}
In summary, HasChor aims to make CP easy to \emph{use}, and it succeeds at that goal --- provided that the user is a Haskell programmer.
But the HasChor design does not make CP easy to \emph{implement} in one's language of choice, and it suffers from efficiency and practicality drawbacks.
Our aim in the rest of this paper is to democratize the \emph{implementation} of library-level choreographic programming while improving its efficiency and practicality.

% !TEX root = ./main.tex

\section{Endpoint Projection as Dependency Injection}
\label{sec:epp-as-di}

A central concept of CP is that a single choreography exhibits different behaviors depending on the location to which it is projected.
Each local computation may or may not be executed, and each communication becomes a send, a receive, or a no-op.
Allowing a caller (in this case, endpoint projection) to modify the behavior of the callee (in this case, a choreography) is a common pattern in software engineering to improve code reusability and testability.
One technique to achieve this is through dependency injection (DI)~\cite{Fowler_2004}.
In DI, the callee receives its dependencies from the caller, who can alter the callee's behavior by providing different dependencies.

In this section, we present endpoint projection as dependency injection (EPP-as-DI), a new technique for implementing library-level choreographic programming.
The key idea of EPP-as-DI is that we can implement CP by representing a choreography as a host-language function that takes \emph{choreographic operators} as arguments.
Then, endpoint projection can change the behavior of the choreography by injecting specialized implementations of the choreographic operators, depending on the projection target.
This technique can be used in any host language that supports higher-order functions, enabling the straightforward implementation of choreographic programming libraries in a wide variety of languages.
We introduce a simple host language as a stand-in for an arbitrary host language in \Cref{subsec:host-lang}, then show how EPP-as-DI is implemented in \Cref{subsec:di}.

\subsection{Choreographies as Host-Language Programs}
\label{subsec:host-lang}

\begin{figure}
    \small
    \begin{align*}
        a               & : \ts{Type}                                                                                          &                                      \\
        l               & : \ts{Location}                                                                                      &                                      \\
        a \at l         & = \ts{Local} \ a + \ts{Remote}                                                                       & \quad \tr{(Located Values)}           \\
        \ts{Unwrap} \ l & = a \at l \rightarrow a                                                                              & \quad \tr{(Unwrap)}                  \\
        \ts{Choreo} \ a & = \ts{Ops} \rightarrow a                                                                             & \quad \tr{(Choreography)}            \\
        \ts{Ops}        & = \ts{Locally} \times \ts{Comm} \times \ts{Bcast}                                                    & \quad \tr{(Choreographic Operators)} \\
        \ts{Locally}    & = \forall a. \ \ (l : \ts{Location}) \rightarrow (\ts{Unwrap} \ l \rightarrow a) \rightarrow a \at l & \quad \tr{(Local Computation)}       \\
        \ts{Comm}       & = \forall a. \ \ (s \ r : \ts{Location}) \rightarrow a \at s \rightarrow a \at r                     & \quad \tr{(Communication)}           \\
        \ts{Bcast}      & = \forall a. \ \ (l : \ts{Location}) \rightarrow a \at l \rightarrow a                               & \quad \tr{(Broadcast)}
    \end{align*}
    \caption{The interface provided by the host-language library for expressing choreographies.}
    \label{fig:choreographies-as-functions-with-dependencies}
\end{figure}

To introduce EPP-as-DI, we assume a simple ML-like host language that supports higher-order functions.
For ease of exposition in this section, our host language is typed; however, types are not essential to implement EPP-as-DI.
Choreographies are expressed as host-language functions using the interface presented in \Cref{fig:choreographies-as-functions-with-dependencies}, which we now describe.

\subsubsection{Located Values}

We assume a set of \ts{Location}s with decidable equality and write them as $l$.
A \emph{located value}, written $a \at l$, is a value of type $a$ at location $l$.
A located value can either be a $\ts{Local} \ a$, meaning the value is at the current location, or a $\ts{Remote}$, meaning the value is at some remote location.
We maintain the invariant that, when doing endpoint projection for $l$, $a \at l$ is always a \ts{Local}.
To use a located value at $l$, it needs to be \emph{unwrapped} first. Since it does not make sense to unwrap a remote value, we provide an $\ts{Unwrap} \ l$ function that can only unwrap values at $l$.  Given a value of type $a \at l$, $\ts{Unwrap} \ l$ produces a value of type $a$.

\subsubsection{Choreographies}

A choreography \ts{Choreo $a$} is a function that takes a set of choreographic operators \ts{Ops} as dependencies and returns some result of type $a$.
The host-language library interface provides three choreographic operators that are sufficient to realize the key elements of CP described in \Cref{subsec:choreographic-approach}.  We will use lower-case \ts{locally}, \ts{comm}, and \ts{bcast} as the names of operators that have types \ts{Locally}, \ts{Comm}, and \ts{Bcast}, respectively:
\begin{itemize}
    \item
          \ts{locally} performs a local computation: it takes a location and a function and runs the function locally at the location.
    \item
          \ts{comm} communicates a value between two locations: it takes a sender and a receiver location, a value at the sender, and returns the same value at the receiver.
    \item
          \ts{bcast} broadcasts a value to the group of locations involved in the interaction: it takes a sender location, a value at the sender, and returns a value at all locations.
\end{itemize}

\begin{figure}
    \small
    \begin{align*}
         & \ts{bookseller} : \ts{Choreo} \ (\ts{Option} \ \ts{Date} \at \ts{buyer})                                                                                                                                       \\
         & \ts{bookseller}(\ts{locally}, \ts{comm}, \ts{bcast}) =                                                                                                                                  \\
         & \quad \tsb{let} \ \ts{title}_\ts{buyer} = \ts{locally}(\ts{buyer}, \lambda(\ts{un}) \rightarrow \ts{input}()) \ \tsb{in}                                                                \\
         & \quad \tsb{let} \ \ts{title}_\ts{seller} = \ts{comm}(\ts{buyer}, \ts{seller}, \ts{title}_\ts{buyer}) \ \tsb{in}                                                                         \\
         & \quad \tsb{let} \ \ts{price}_\ts{seller} = \ts{locally}(\ts{seller}, \lambda(\ts{un}) \rightarrow {\ts{catalog}.\ts{get\_price}}(\ts{un}(\ts{title}_\ts{seller}))) \ \tsb{in}           \\
         & \quad \tsb{let} \ \ts{price}_\ts{buyer} = \ts{comm}(\ts{seller}, \ts{buyer}, \ts{price}_\ts{seller}) \ \tsb{in}                                                                         \\
         & \quad \tsb{let} \ \ts{decision}_\ts{buyer} = \ts{locally}(\ts{buyer}, \lambda(\ts{un}) \rightarrow \ts{un}(\ts{price}_\ts{buyer}) \leq \ts{budget}) \ \tsb{in}                          \\
         & \quad \tsb{let} \ \ts{decision} = \ts{bcast}(\ts{buyer}, \ts{decision}_\ts{buyer}) \ \tsb{in}                                                                                           \\
         & \quad \tsb{if} \ \ts{decision} \ \tsb{then}                                                                                                                                             \\
         & \quad \quad \tsb{let} \ \ts{delivery}_\ts{seller} = \ts{locally}(\ts{seller}, \lambda(\ts{un}) \rightarrow \ts{catalog}.\ts{get\_delivery}(\ts{un}(\ts{title}_\ts{seller}))) \ \tsb{in} \\
         & \quad \quad \tsb{let} \ \ts{delivery}_\ts{buyer} = \ts{comm}(\ts{seller}, \ts{buyer}, \ts{delivery}_\ts{seller}) \ \tsb{in}                                                             \\
         & \quad \quad \ts{locally}(\ts{buyer}, \lambda(\ts{un}) \rightarrow \ts{Some}(\ts{un}(\ts{delivery}_\ts{buyer}))) \\
         & \quad \tsb{else} \\
         & \quad \quad \ts{locally}(\ts{buyer}, \lambda(\ts{un}) \rightarrow \ts{None})
    \end{align*}
    \caption{Bookseller Choreography}
    \label{fig:bookseller-function}
\end{figure}

We can write choreographies as functions of type \ts{Choreo $a$} by using the provided choreographic operators in the body of the function.
To illustrate, \Cref{fig:bookseller-function} shows the bookseller protocol implemented in our notional host language using the API of \Cref{fig:choreographies-as-functions-with-dependencies}.
We assume that the host language supports standard language constructs such as \tsb{let ... in} and \tsb{if ... then ... else}.
The bookseller choreography uses \ts{bcast} to propagate knowledge of choice and implement conditionals.
When the buyer makes a decision (${\ts{decision}_\ts{buyer}}$), it is broadcasted to all locations (\ts{decision}).
Since all locations have the same data, it is safe to use the control-flow constructs of the host language, such as \tsb{if}, to implement conditionals in choreographies.

\subsection{Endpoint Projection as Injecting Dependencies}
\label{subsec:di}

\begin{figure}
    \small
    \begin{align*}
         & \ts{epp} : \ts{Choreo} \ a \rightarrow [\ts{Location}] \rightarrow \ts{Location} \rightarrow a                                                                                 \\
         & \ts{epp}(c, \ti{ls}, l) =                                                                                                                                                      \\
         & \quad \tsb{let} \ \ts{unwrap}(v) = \tsb{if} \ \tsb{let} \ \ts{Local}(a) = v \ \tsb{then} \ a \ \tsb{else} \ \ts{error}(\ts{``impossible''}) \ \tsb{in}                           \\
         & \quad \tsb{let} \ \ts{locally}(l', f) = \tsb{if} \ l == l' \ \tsb{then} \ \ts{Local}(f(\ts{unwrap})) \ \tsb{else} \ \ts{Remote} \ \tsb{in}                                     \\
         & \quad \tsb{let} \ \ts{comm}(s, r, a) =                                                                                                                                         \\
         & \quad \quad \tsb{if} \ l == s \ \tsb{then} \ \ts{send}(\ts{unwrap}(a), r);\ts{Remote} \ \tsb{else if} \ l == r \ \tsb{then} \ \ts{Local}(\ts{recv}(s)) \ \tsb{else} \ \ts{Remote} \ \tsb{in} \\
         & \quad \tsb{let} \ \ts{bcast}(s, a) = \tsb{if} \ l == s \ \tsb{then} \ \forall r \in \ti{ls}.\, \ts{send}(\ts{unwrap}(a), r) ; \ts{unwrap}(a) \ \tsb{else} \ \ts{recv}(s) \ \tsb{in}          \\
         & \quad c(\ts{locally}, \ts{comm}, \ts{bcast})
    \end{align*}
    \caption{Endpoint Projection as Injecting Dependencies}
    \label{fig:endpoint-projection-as-injecting-dependencies}
\end{figure}

Since a choreography is a function that takes choreographic operators as dependencies, we can determine the meaning of these operators by injecting specialized implementations of them, leading to the definition of endpoint projection as a host-language function $\ts{epp}$, shown in \Cref{fig:endpoint-projection-as-injecting-dependencies}.
We assume the existence of $\ts{send}$ and $\ts{recv}$ functions in the host language that implement message transport, for instance, by calling into a host-language networking library.
$\ts{epp}$ takes a choreography $c$, a list of locations participating in the choreography $\ti{ls}$, and a target location $l$, then projects the choreography to a node-local program for the target location.
Inside $\ts{epp}$, we construct the three choreographic operators from the viewpoint of $l$ and supply them to $c$:
\begin{itemize}
    \item
          For operator $\ts{locally}(l', f)$, if $l$ is the same as $l'$, we perform the local computation $f$; otherwise, no action is taken.
    \item
          For operator $\ts{comm}(s, r, a)$, if $l$ is the same as the sender location $s$, we perform a $\ts{send}$ of $a$ to the receiver;
          or if $l$ is the same as the receiver location $r$, we perform a $\ts{recv}$ from the sender;
          otherwise, no action is taken.
    \item
          For operator $\ts{bcast}(s, a)$, if $l$ is the same as the sender location $s$, we perform a series of $\ts{send}$s of $a$ to all the locations participating in the interaction; otherwise, no action is taken.
\end{itemize}

We used the EPP-as-DI technique to implement ChoRus, a choreographic programming library for Rust. We describe the design and implementation of ChoRus in \Cref{sec:chorus}.

% !TEX root = ./main.tex

\section{Efficient Conditionals with Choreographic Enclaves}
\label{sec:enclaves}

As discussed in \Cref{subsec:choreographic-approach}, implementing conditionals in choreographic programming is challenging because of the ``knowledge of choice'' problem~\citep{castagna-knowledge-of-choice}. A CP language must ensure --- either statically or dynamically --- that choreographies propagate knowledge of the outcome of evaluating a conditional expression to all locations that are affected by the choice. If CP is implemented as a standalone language, then the compiler can perform static analysis to check this property,
and a choreography that fails to propagate knowledge of choice is deemed \emph{unprojectable}.
Standalone CP languages can even support choreography \emph{amendment}~\citep{cruz-filipe-montesi-core, lanese-amending-choreographies, basu-choreography-repair,cruz-filipe-certified-repair}, a procedure that determines if a choreography is unprojectable as-is and then automatically inserts the minimum necessary communication to make it projectable.

Without access to the full AST of programs in the CP language, however, static analysis becomes infeasible.
In particular, with both the EPP-as-DI approach of \Cref{sec:epp-as-di} and in HasChor's freer-monad-based approach, we cannot perform static analysis on choreographies to determine how knowledge of choice needs to be propagated.
With static analysis off the table as an option, then propagation of knowledge of choice needs to be handled some other way.
In \Cref{sec:epp-as-di}, we solved the problem in a naive way by implementing conditionals with broadcast, which ensures that \emph{all} locations receive the knowledge of choice, whether they are affected by the choice or not. HasChor's \hs{cond} operator internally uses broadcast as well. Not only does this naive approach introduce unnecessary communication,
it may cause an undesired leak of information to locations who should not have it.

Alternatively, we could do without static analysis another way: by requiring the programmer to provide annotations to convey their intent.  In fact, in the absence of choreography amendment, this is the typical approach even in standalone CP languages: the programmer must annotate the branches of a conditional with selection annotations that indicate to the compiler that knowledge of choice must be propagated, as we see in the Choral code in \Cref{fig:bookseller-choral} that uses the \code{select} method.
Yet the approach of adding selection annotations is somewhat unsatisfying, because we must add annotations to make our code \emph{correct} (that is, projectable).  If we must annotate our code for the benefit of the compiler, it would be preferable if we could begin with a choreography that is \emph{correct, but inefficient}, and then add annotations to make it \emph{efficient}.

In this section, we address this design challenge with \emph{choreographic enclaves}, a novel CP language feature. Enclaves are sub-choreographies that execute at a specified subset of the locations involved in a larger choreography.  One may broadcast within an enclave, just like in any other choreography, but the broadcast will only go to those locations that are in the specified subset. Enclaves allow finer control over the propagation of knowledge of choice, enabling an efficient implementation of conditionals in library-level CP without static analysis.

In \Cref{subsec:two-buyer}, we present a variant of the bookseller protocol to motivate the need for fine-grained control over propagation of knowledge of choice. Then, in \Cref{subsec:enclave-op} we introduce choreographic enclaves. We define an \ts{enclave} operator and present its type signature and implementation, and we show how to implement the two-buyer protocol with \ts{enclave} and compare it with the naive approach and the selection-annotation approach.

\subsection{The Two-Buyer Protocol}
\label{subsec:two-buyer}

To illustrate the problem of inefficient conditionals, let us consider a variant of the bookseller protocol: the \emph{two-buyer} protocol~\cite{honda-mpsts, hirsch2022pirouette}.
In this protocol, there are two buyers who wish to collectively buy a book from the seller.
First, \ts{buyer1} sends the title to the seller, and the seller sends the price to both buyers.
Then, \ts{buyer2} tells \ts{buyer1} how much they can contribute, and \ts{buyer1} decides whether to buy the book by comparing the price with the buyers' combined budget.
If \ts{buyer1} decides to buy the book, they send their intent to buy to the seller, and the seller sends the delivery date to \ts{buyer1}.
Otherwise, \ts{buyer1} tells the seller that they will not buy the book.

\begin{figure}
    \small
    \begin{align*}
         & \ts{two\_buyer} : \ts{Choreo} \ (\ts{Option} \ \ts{Date} \at \ts{buyer1}) \\
         & \ts{two\_buyer}(\ts{locally}, \ts{comm}, \ts{bcast}) = \\
         & \quad \tsb{let} \ \ts{title}_\ts{buyer1} = \ts{locally}(\ts{buyer1}, \lambda(\ts{un}) \rightarrow \ts{input}()) \ \tsb{in} \\
         & \quad \tsb{let} \ \ts{title}_\ts{seller} = \ts{comm}(\ts{buyer1}, \ts{seller}, \ts{title}_\ts{buyer1}) \ \tsb{in} \\
         & \quad \tsb{let} \ \ts{price}_\ts{seller} = \ts{locally}(\ts{seller}, \lambda(\ts{un}) \rightarrow \ts{catalog}.\ts{get\_price}(\ts{un}(\ts{title}_\ts{seller}))) \ \tsb{in} \\
         & \quad \tsb{let} \ \ts{price}_\ts{buyer1} = \ts{comm}(\ts{seller}, \ts{buyer1}, \ts{price}_\ts{seller}) \ \tsb{in} \\
         & \quad \tsb{let} \ \ts{price}_\ts{buyer2} = \ts{comm}(\ts{seller}, \ts{buyer2}, \ts{price}_\ts{seller}) \ \tsb{in} \\
         & \quad \tsb{let} \ \ts{contribution} = \ts{comm}(\ts{buyer2}, \ts{buyer1}, \ts{buyer2\_budget}) \ \tsb{in} \\
         & \quad \tsb{let} \ \ts{decision}_\ts{buyer1} = \ts{locally}(\ts{buyer1}, \lambda(\ts{un}) \rightarrow \ts{un}(\ts{price}_\ts{buyer1}) \leq \ts{buyer1\_budget} + \ts{contribution}) \ \tsb{in} \\
         & \quad \tsb{let} \ \ts{decision} = \ts{bcast}(\ts{buyer1}, \ts{decision}_\ts{buyer1}) \ \tsb{in} \\
         & \quad \tsb{if} \ \ts{decision} \ \tsb{then} \\
         & \quad \quad \tsb{let} \ \ts{delivery}_\ts{seller} = \ts{locally}(\ts{seller}, \lambda(\ts{un}) \rightarrow \ts{catalog}.\ts{get\_delivery}(\ts{un}(\ts{title}_\ts{seller}))) \ \tsb{in} \\
         & \quad \quad \tsb{let} \ \ts{delivery}_\ts{buyer1} = \ts{comm}(\ts{seller}, \ts{buyer1}, \ts{delivery}_\ts{seller}) \ \tsb{in} \\
         & \quad \quad \ts{locally}(\ts{buyer1}, \lambda(\ts{un}) \rightarrow \ts{Some}(\ts{un}(\ts{delivery}_\ts{buyer1}))) \\
         & \quad \tsb{else} \\
         & \quad \quad \ts{locally}(\ts{buyer1}, \lambda(\ts{un}) \rightarrow \ts{None})
    \end{align*}
    \caption{A naive version of the two-buyer protocol with \ts{bcast}}
    \label{fig:two-buyer-function-bcast}
\end{figure}

\begin{figure}
    \centering
    \begin{subfigure}[c]{.5\textwidth}
        \centering
        \includegraphics[width=5cm]{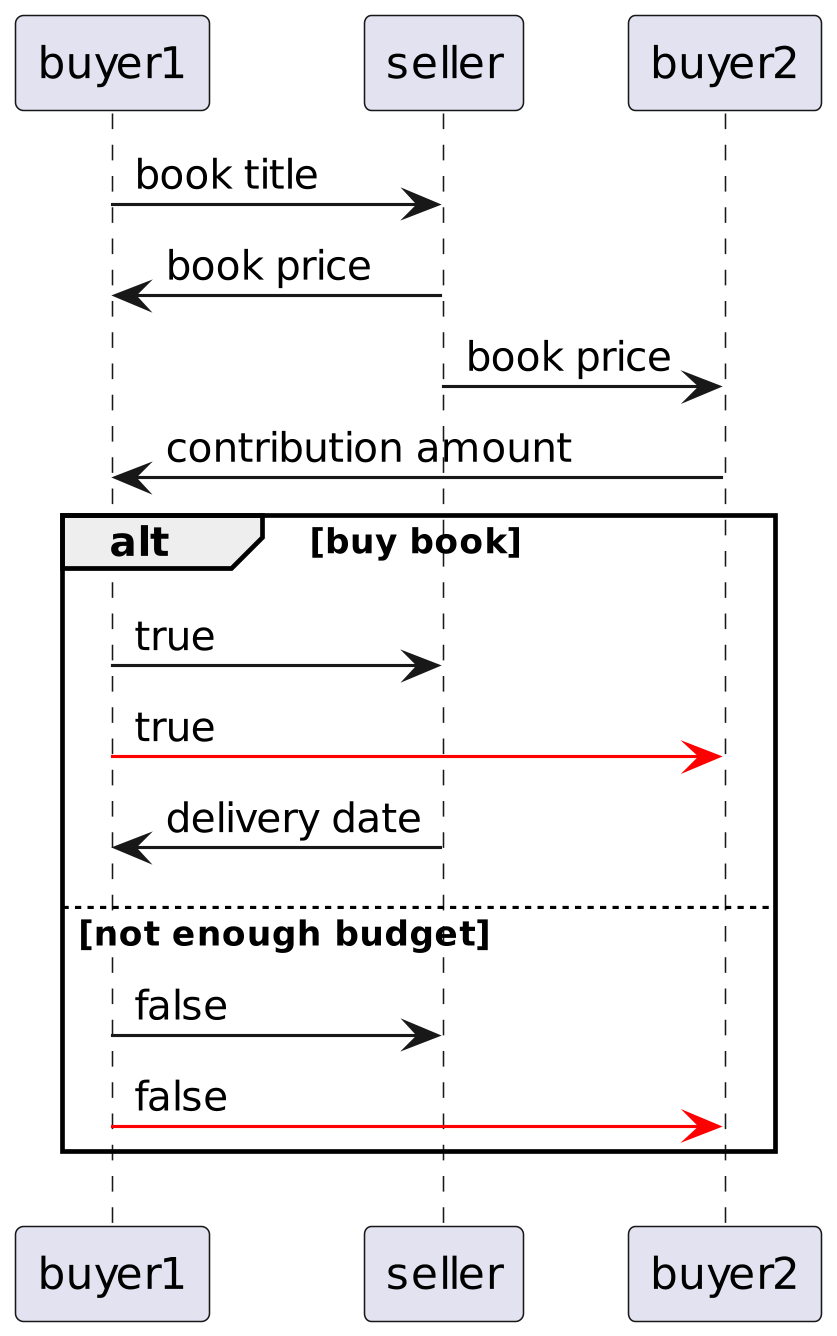}
        \caption{Using broadcast naively}
        \label{fig:two-buyer-broadcast-sequence}
    \end{subfigure}%
    \begin{subfigure}[c]{.5\textwidth}
        \centering
        \includegraphics[width=5cm]{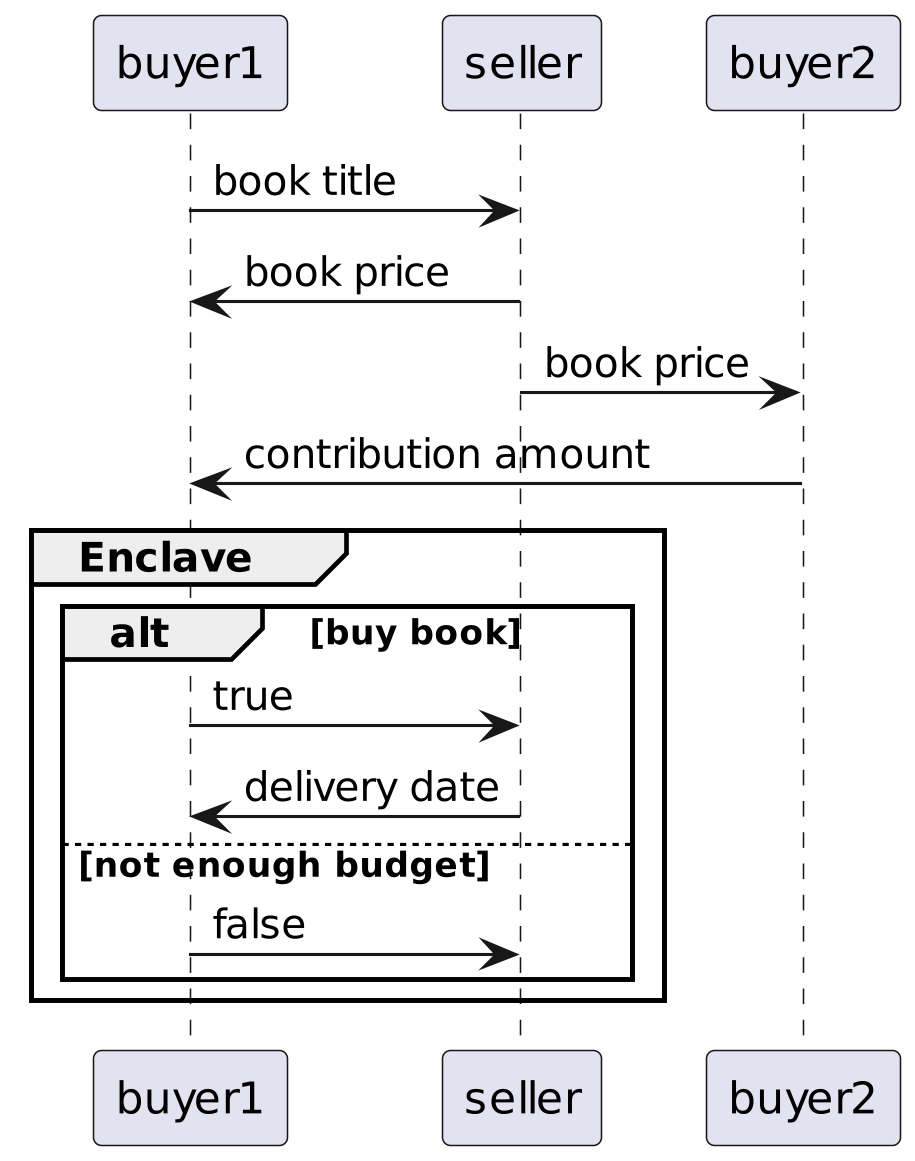}
        \caption{With enclave}
        \label{fig:two-buyer-enclave-sequence}
    \end{subfigure}
    \caption{Sequence diagrams of the two-buyer protocol with and without enclave}
    \Description{Sequence diagrams of the two-buyer protocol with and without enclave}
    \label{fig:two-buyer-seq}
\end{figure}

Using the \ts{bcast} choreographic operator that we introduced in \Cref{subsec:host-lang}, we can implement the two-buyer protocol in our notional host language, as shown in \Cref{fig:two-buyer-function-bcast}. \Cref{fig:two-buyer-broadcast-sequence} shows a sequence diagram of the execution of the protocol.
After \ts{buyer1} makes a decision, to perform the conditional, it broadcasts the decision to all locations, i.e., the seller and \ts{buyer2}.
It is important that the seller receives the decision because the seller needs to know whether to send a delivery date to \ts{buyer1}, but \ts{buyer2} does not need to receive the decision, as its subsequent behavior does not depend on it.
Nonetheless, because of broadcast, \ts{buyer2} receives the decision, causing unnecessary communication between \ts{buyer1} and \ts{buyer2}, shown in red in \Cref{fig:two-buyer-broadcast-sequence}.
While this communication does not affect the correctness of the choreography, it is inefficient and can be problematic in more complex choreographies with many participants.
Moreover, it leaks information about choice, which can be a security concern.
For example, \ts{buyer2} might infer the budget of \ts{buyer1} by observing the decision, and this type of information leakage might be undesirable in some applications.

\subsection{The Enclave Operator}
\label{subsec:enclave-op}

To prevent unnecessary communication, we introduce the \ts{enclave} choreographic operator. The \ts{enclave} operator executes a sub-choreography at a specified set of locations. Inside the sub-choreography, the \ts{broadcast} operator sends data only to locations in the specified set. This allows us to perform conditionals without sending data to unaffected locations.

We extend the interface of our host-language library from \Cref{fig:choreographies-as-functions-with-dependencies} to add support for an \ts{enclave} operator with the type \ts{Enclave}, specified below:
{\small
\begin{align*}
    \ts{Ops}       & = \ts{Locally} \times \ts{Comm} \times \ts{Bcast} \times \ts{Enclave}                 & \quad \textrm{(Choreographic Operators)} \\
    \ts{Enclave} & = \forall a, l. \ [\ts{Location}] \rightarrow \ts{Choreo} \ a \at l \rightarrow a \at l & \quad \textrm{(Enclave)}
\end{align*}}
The first argument to \ts{enclave} is a list of locations where the sub-choreography is to be executed, and the second argument is the sub-choreography. It returns the result of running the sub-choreography.
To implement endpoint projection for \ts{enclave}, we update the definition of \ts{epp} from \Cref{fig:endpoint-projection-as-injecting-dependencies} as follows:
{\small
\begin{align*}
  & \ts{epp} : \ts{Choreo} \ a \rightarrow [\ts{Location}] \rightarrow \ts{Location} \rightarrow a \\
  & \ts{epp}(c, \ti{ls}, l) = \\
  & \quad ... \\
  & \quad \tsb{let} \ \ts{enclave}({ls}', c') = \tsb{if} \ l \in {ls}' \ \tsb{then} \ \ts{epp}(c', {ls}', l) \ \tsb{else} \ \ts{Remote} \ \tsb{in} \\
  & \quad c(\ts{locally}, \ts{comm}, \ts{bcast}, \ts{enclave})
\end{align*}}
The \ts{enclave} operator recursively calls the sub-choreography by calling \ts{epp} with the sub-choreography and the list of locations where the sub-choreography is executed if the projection target is one of the specified locations. The behavior of \ts{bcast} inside the sub-choreography depends on the ${ls}'$ argument to the recursive call to \ts{epp}, so \ts{bcast} inside the sub-choreography will only send data to the specified locations.
\begin{figure}
    \small
    \begin{align*}
         & \ts{two\_buyer} : \ts{Choreo} \ (\ts{Option} \ \ts{Date} \at \ts{buyer1}) \\
         & \ts{two\_buyer}(\ts{locally}, \ts{comm}, \ts{bcast}, \ts{enclave}) = \\
         & \quad ... \\
         & \quad \tsb{let} \ \ts{decision}_\ts{buyer1} = \ts{locally}(\ts{buyer1}, \lambda(\ts{un}) \rightarrow \ts{un}(\ts{price}_\ts{buyer1}) \leq \ts{buyer1\_budget} + \ts{contribution}) \ \tsb{in} \\
         & \quad \tsb{let} \ c(\ts{locally}, \ts{comm}, \ts{bcast}, \ts{enclave}) = \\
         & \quad \quad \tsb{let} \ \ts{decision} = \ts{bcast}(\ts{buyer1}, \ts{decision}_\ts{buyer1}) \ \tsb{in} \\
         & \quad \quad \tsb{if} \ \ts{decision} \ \tsb{then} \\
         & \quad \quad \quad \tsb{let} \ \ts{delivery}_\ts{seller} = \ts{locally}(\ts{seller}, \lambda(\ts{un}) \rightarrow \ts{catalog}.\ts{get\_delivery}(\ts{un}(\ts{title}_\ts{seller}))) \ \tsb{in} \\
         & \quad \quad \quad \tsb{let} \ \ts{delivery}_\ts{buyer1} = \ts{comm}(\ts{seller}, \ts{buyer1}, \ts{delivery}_\ts{seller}) \ \tsb{in} \\
         & \quad \quad \quad \ts{locally}(\ts{buyer1}, \lambda(\ts{un}) \rightarrow \ts{Some}(\ts{un}(\ts{delivery}_\ts{buyer1}))) \\
         & \quad \quad \tsb{else} \\
         & \quad \quad \quad \ts{locally}(\ts{buyer1}, \lambda(\ts{un}) \rightarrow \ts{None}) \\
         & \quad \textbf{in} \\
         & \quad \ts{enclave}([\ts{buyer1}, \ts{seller}], c)
    \end{align*}
    \caption{A more efficient version of the two-buyer protocol using an enclave}
    \label{fig:two-buyer-function-enclave}
\end{figure}
Using \ts{enclave}, we can rewrite the last part of the two-buyer protocol, as shown in \Cref{fig:two-buyer-function-enclave}.
After \ts{buyer1} makes a decision, we define a sub-choreography $c$ that uses \ts{bcast} to perform conditionals. Then, we call the sub-choreography at \ts{buyer1} and seller using \ts{enclave}. Because the sub-choreography is not executed at \ts{buyer2}, \ts{bcast} does not send the decision to \ts{buyer2}, as shown in \Cref{fig:two-buyer-enclave-sequence}.

While we have shown how to implement endpoint projection for \ts{enclave} using the EPP-as-DI technique, the use of choreographic enclaves is orthogonal to the use of EPP-as-DI.  For instance, one could extend HasChor with an \ts{enclave} operator without departing from HasChor's freer-monad-based implementation of EPP.

% !TEX root = ./main.tex

\section{ChoRus: Library-Level Choreographic Programming for Rust}
\label{sec:chorus}

In this section, we present ChoRus, the first choreographic programming library for the Rust programming language. ChoRus is implemented using EPP-as-DI, supports choreographic enclaves, and has other features that make it a practical choice for distributed programming in Rust. We describe how we encode EPP-as-DI in Rust~\Cref{subsec:epp-as-di-chorus}, and give a brief tour of ChoRus features~\Cref{subsec:chorus-features}.  The code shown in this section is simplified for presentational purposes. ChoRus is open source, and its implementation, case studies and benchmarking code, and documentation are available at \url{https://github.com/lsd-ucsc/ChoRus}.

\subsection{EPP-as-DI in ChoRus}
\label{subsec:epp-as-di-chorus}

\subsubsection{Locations}

ChoRus represents each location at which node-local code runs as a distinct type. In Rust, we can create a new type by defining a struct. Locations must be comparable for equality to perform endpoint projection. To that end, ChoRus defines the \rs{ChoreographyLocation} trait, which all location types must implement:
\begin{minted}{rust}
trait ChoreographyLocation: Copy {
    fn name() -> &'static str;
}
\end{minted}
The \rs{name} method returns the string representation of the location, which is used to compare locations for equality. Thanks to Rust's macro system, \rs{ChoreographyLocation} can be derived automatically. For example, the following code defines a location named \rs{Alice}:

\begin{minted}{rust}
#[derive(ChoreographyLocation)]
struct Alice;
\end{minted}

\subsubsection{Located Values}

Located values are values that reside at a specific location. ChoRus defines the \rs{Located<V, L1>} struct to represent a located value of type \rs{V} at location \rs{L1}:

\begin{minted}{rust}
struct Located<V, L1: ChoreographyLocation> {
    value: Option<V>,
    phantom: PhantomData<L1>,
}
\end{minted}

The \rs{value} field holds a value of type \rs{Option<V>}; it is \rs{Some} if the current projection target is \rs{L1} and \rs{None} otherwise. We use \rs{std::marker::PhantomData} to indicate to the compiler that the \rs{L1} parameter is not used at run time.

\subsubsection{Choreography Trait}

In \Cref{sec:epp-as-di}, we represented choreographies as functions. To provide a more ergonomic API, ChoRus represents choreographies as structs that implement the \rs{Choreography} trait. The \rs{Choreography} trait is defined as follows:

\begin{minted}{rust}
trait Choreography<R = ()> {
    fn run(self, op: &impl ChoreoOp) -> R;
}
\end{minted}

The \rs{R} type parameter represents the return type of the choreography. The \rs{run} method takes a reference to an object that implements the \rs{ChoreoOp} trait, which provides the choreographic operators.

\subsubsection{ChoreoOp Trait}

ChoRus supports the four choreographic operators \rs{locally}, \rs{comm}, \rs{broadcast}, and \rs{enclave}, as described in \Cref{sec:epp-as-di} and \Cref{sec:enclaves}.

\Cref{fig:chorus-choreoop} shows an excerpt of the \rs{ChoreoOp} trait that implements the \rs{locally} operator. The \rs{locally} method takes a location \rs{location} and a function \rs{computation} and returns a \rs{Located} value. The \rs{computation} function takes an argument of type \rs{Unwrapper<L1>}, which it can use to unwrap located values at location \rs{L1}. Other choreographic operators are defined similarly as methods of the \rs{ChoreoOp} trait.

\begin{figure}
    \begin{minted}{rust}
trait ChoreoOp {
    fn locally<V, L1: ChoreographyLocation>(
        &self,
        location: L1,
        computation: impl Fn(Unwrapper<L1>) -> V,
    ) -> Located<V, L1>;
    // ...
}
\end{minted}
    \caption{The \rs{ChoreoOp} trait (excerpt).}
    \label{fig:chorus-choreoop}
    \Description{The ChoreoOp trait (excerpt).}
\end{figure}

\subsubsection{Transport}

The \rs{Transport} trait represents the message transport layer. Users can implement the \rs{Transport} trait by providing the \rs{send} and \rs{receive} methods. ChoRus has two built-in transport implementations: \rs{LocalTransport} and \rs{HttpTransport}. The \rs{LocalTransport} implementation models each location as a thread and uses an inter-thread channel to send messages. The \rs{HttpTransport} implementation uses HTTP to send messages.

\subsubsection{Endpoint Projection}

ChoRus provides the \rs{Projector} struct to perform endpoint projection and execute choreographies. First, users construct a \rs{Projector} by passing the projection target and the transport. Then, they can call the \rs{epp_and_run} method to perform endpoint projection and execute the choreography. The \rs{epp_and_run} method takes a choreography, defines \rs{EppOp} --- an object that implements \rs{ChoreoOp} for the projection target --- and calls the \rs{run} method of the choreography with it. \Cref{fig:chorus-projector} shows an excerpt of the \rs{epp_and_run} method and the implementation of \rs{locally}.

\begin{figure}
    \begin{minted}{rust}
impl<...> Projector<...> {
    pub fn epp_and_run<...>(&'a self, choreo: C) -> V {
        struct EppOp<...> {...}
        impl<...> ChoreoOp for EppOp<...>
        {
            fn locally<V, L1: ChoreographyLocation>(
                &self,
                location: L1,
                computation: impl Fn(Unwrapper<L1>) -> V,
            ) -> Located<V, L1> {
                if L1::name() == Target::name() {
                    let value = computation(Unwrapper::new());
                    Located::local(value)
                } else {
                    Located::remote()
                }
            }
            // ...
        }
        choreo.run(&EppOp {...})
    }
}
\end{minted}
    \caption{The \rs{epp_and_run} method of the \rs{Projector} struct.}
    \label{fig:chorus-projector}
    \Description{The epp\_and\_run method of the Projector struct.}
\end{figure}

\subsection{Advanced Features}
\label{subsec:chorus-features}

ChoRus supports all the features supported by HasChor~\citep{HasChor}, such as swappable transport backends, higher-order choreographies, and location polymorphism. In this section, we present two new features of ChoRus: \emph{location sets} and \emph{located input/output}.

\subsubsection{Location Sets}

In ChoRus, the set of locations at which a choreography runs is represented at the type level.
We call this type the \textit{location set} of the choreography.
Each choreography has an associated type \rs{L} that represents its location set.
\rs{ChoreoOp} is parametrized by the location set of the choreography and prevents users from using locations that are not in the location set.
For example, the following code defines a choreography \rs{AliceBobChoreography} that runs on locations \rs{Alice} and \rs{Bob}. \rs{LocationSet!} is a macro that constructs a special location set type. Inside the \rs{run} method, we can only use locations \rs{Alice} and \rs{Bob}. If we try to use location \rs{Carol}, the Rust compiler will report an error.
Location sets are especially useful when defining choreographic enclaves, as they prevent us from accidentally using locations outside the enclave.

\begin{figure}
    \begin{minted}{rust}
struct AliceBobChoreography;
impl Choreography for AliceBobChoreography {
    type L = LocationSet!(Alice, Bob);
    fn run(self, op: &impl ChoreoOp<Self::L>) {
        op.locally(Carol, |_| println!("Hello from Carol!"));
    }
}
\end{minted}
    \caption{Invalid use of location \rs{Carol} in \rs{AliceBobChoreography}.}
    \label{fig:chorus-location-set-example}
    \Description{Invalid use of location Carol in AliceBobChoreography.}
\end{figure}

\subsubsection{Located Input/Output}
\label{subsec:chorus-located-io}

When a choreography is used as part of a larger program, it is often useful to be able to pass \emph{located} values to and from the choreography.
For example, consider a simple password authentication protocol between a client and a server. The client reads a password from the user and sends it to the server. The server checks the password and sends the result back to the client. The client then prints the result. The inputs to this choreography are (1) the typed password on the client, and (2) the correct password on the server, and the output is the result of the authentication on the client.
Morally, these are all \emph{located} values; for example, when running the choreography on the client, we do not have access to the correct password on the server.
However, from outside the choreography, we do not have a way to talk about their locations.
To solve this problem, ChoRus provides a \emph{located input/output} feature that provides a convenient and type-safe way to handle located values.

\rs{Projector} plays an important role in the located input/output feature. \rs{Projector} is parameterized by the projection target and can construct (1) local located values at the projection target, and (2) remote located values at other locations. It can also unwrap located values at the projection target, but not at other locations.

\Cref{fig:chorus-located-io-example} shows the password authentication choreography and code to execute the choreography as the client and as the server. When running the choreography as the client, we use an instance of \rs{Projector} that is parameterized by the client location. We provide the password attempt as a local located value and the correct password as a remote located value on the server. Conversely, when running the choreography as the server, we provide the password attempt as a remote value and the correct password as a local value. The result can only be unwrapped at the client location using the \rs{unwrap} method of \rs{Projector}.

\begin{figure}
    \begin{subfigure}{\textwidth}
        \begin{minted}{rust}
struct PasswordAuthChoreography {
    attempt_password: Located<String, Client>,
    correct_password: Located<String, Server>,
}
impl Choreography<Located<bool, Client>> for PasswordAuthChoreography {
    type L = LocationSet!(Client, Server);
    fn run(self, op: &impl ChoreoOp<Self::L>) -> Located<bool, Client> {
        let password = op.comm(Client, Server, &self.attempt_password);
        let result = op.locally(Server, |un| {
            un.unwrap(&password) == un.unwrap(&self.correct_password)
        });
        op.comm(Server, Client, &result)
    }
}
        \end{minted}
        \caption{Password authentication choreography}
        \vspace{1em}
    \end{subfigure}
    \begin{subfigure}{\textwidth}
        \begin{minted}{rust}
let result = client_projector.epp_and_run(PasswordAuthChoreography {
    attempt_password: client_projector.local("1234".to_string()),
    correct_password: client_projector.remote(Server),
});
println!("Result: {}", client_projector.unwrap(result));
                    \end{minted}
        \caption{Client code}
        \vspace{1em}
    \end{subfigure}
    \begin{subfigure}{\textwidth}
        \begin{minted}{rust}
server_projector.epp_and_run(PasswordAuthChoreography {
    attempt_password: server_projector.remote(Client),
    correct_password: server_projector.local("password".to_string()),
});
        \end{minted}
        \caption{Server code}
    \end{subfigure}
    \caption{The password authentication choreography (a), along with node-local code to invoke it on the client (b) and server (c).}
    \label{fig:chorus-located-io-example}
    \Description{The password authentication choreography (a), along with node-local code to invoke it on the client (b) and server (c)}
\end{figure}

% !TEX root = ./main.tex

\section{Evaluation}
\label{sec:evaluation}

In this section, we assess the utility and practicality of library-level CP with ChoRus.
First, to demonstrate that ChoRus indeed brings the advantages of CP to Rust, we present a case study involving a key-value store~(\Cref{subsec:eval-kvs}). In this case study, we implement a simple replicated key-value store as a choreography and as a traditional Rust program. We compare these two implementations and highlight how choreography helps to track the flow of data and control.
Next, to illustrate that library-level CP enables code reuse, we conduct a second case study: a multiplayer tic-tac-toe game~(\Cref{subsec:eval-ttt}).
We begin by showing the code for a tic-tac-toe game that runs locally, then we use ChoRus to modify the program to run across multiple computers over a network with minimal changes. We observe that library-level CP allows a substantial portion of the local code to be reused for the distributed implementation.
Finally, we measure the performance overhead incurred by using ChoRus~(\Cref{subsec:eval-perf}). Through benchmarking, we show that ChoRus introduces very minimal overhead, making it sufficiently practical for use.

\subsection{Case Study 1: Replicated Key-Value Store}
\label{subsec:eval-kvs}

To demonstrate how ChoRus helps developers to implement distributed systems, we consider a simple replicated key-value store. Our key-value store supports two operations: \rs{get} and \rs{put}. The \rs{get} operation takes a key and returns the value associated with the key. The \rs{put} operation takes a key and a value, and associates the key with the value. Our system consists of three nodes: \rs{Client}, \rs{Primary}, and \rs{Backup}. The \rs{Client} node takes a request from the user and sends the request to the \rs{Primary} node. The \rs{Primary} node checks the type of the request. If the request is a \rs{get} request, it looks up the requested key in its local state and returns the response to the client. If the request is a \rs{put} request, it forwards the request to the backup node. The backup node updates its local state and returns the response to the \rs{Primary} node. Once \rs{Primary} receives the response from \rs{Backup}, it applies the update to its local state and returns the response to the client.

\begin{figure}
    \centering
    \begin{subfigure}[c]{0.49\textwidth}
        \centering
        \includegraphics[width=0.95\textwidth]{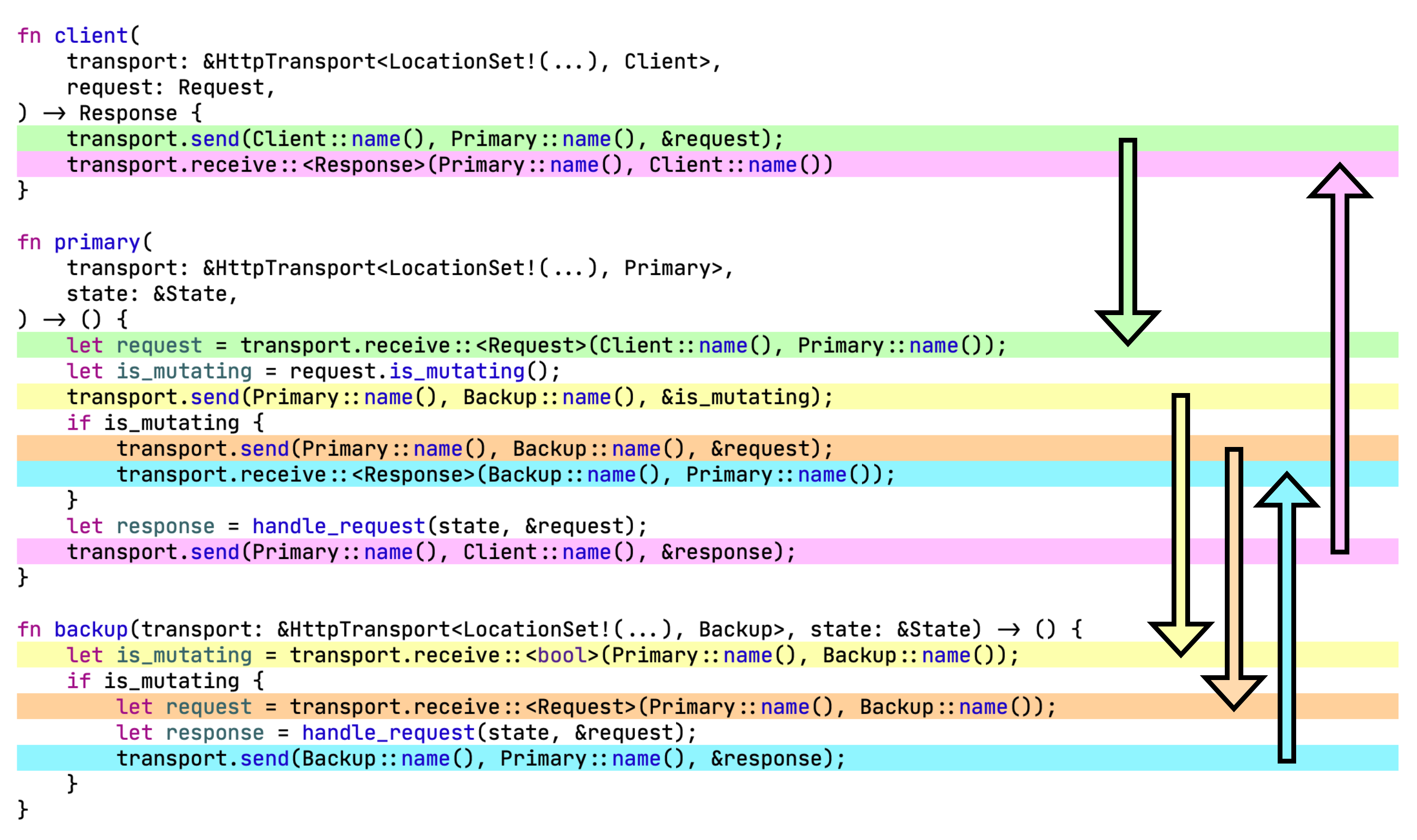}
        \caption{Handwritten Rust KVS}
        \Description{Handwritten Rust KVS}
        \label{fig:kvs-handwritten-flow}
    \end{subfigure}
    \begin{subfigure}[c]{0.49\textwidth}
        \centering
        \includegraphics[width=\textwidth]{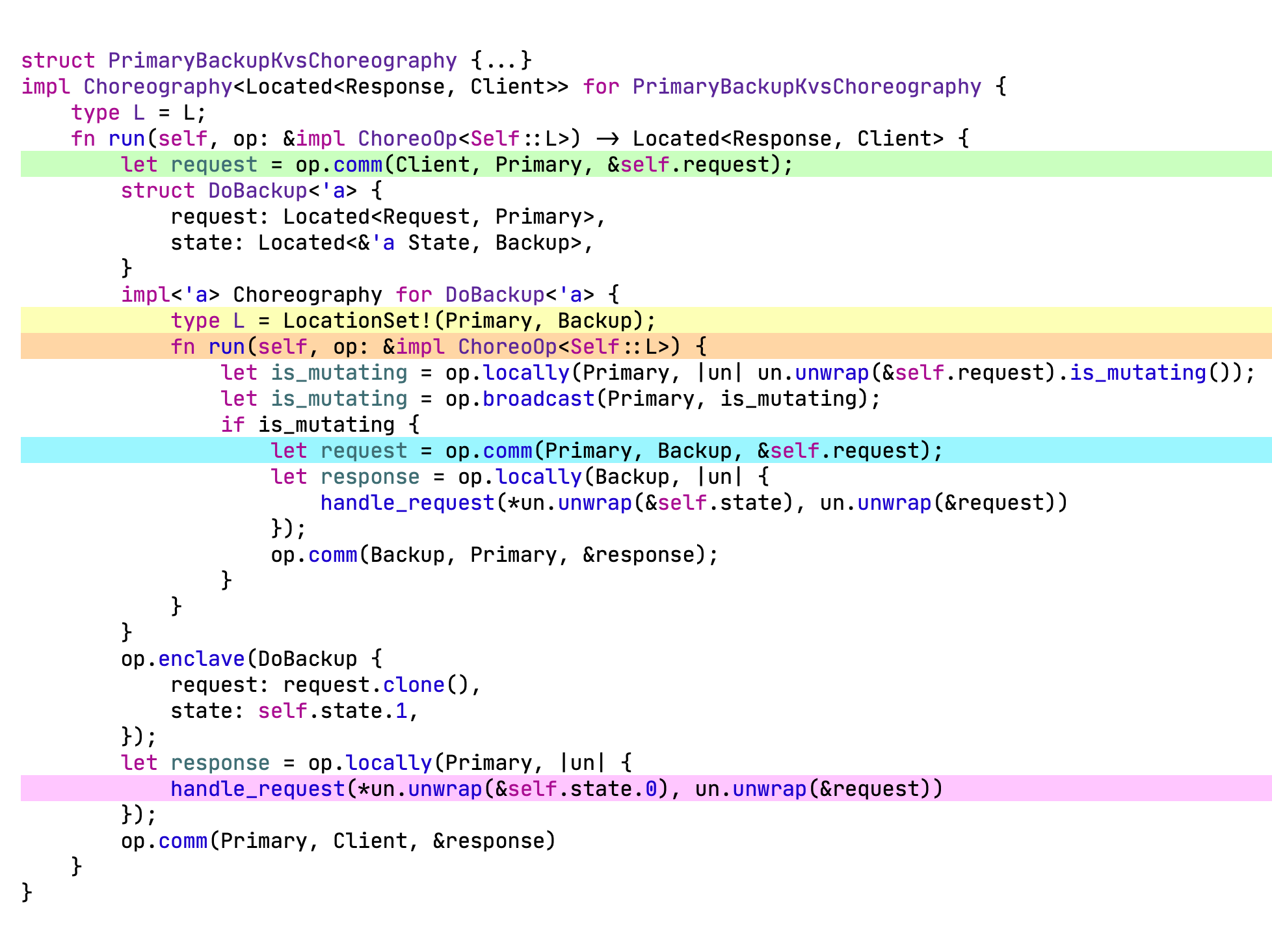}
        \caption{Choreographic ChoRus KVS}
        \Description{Choreographic ChoRus KVS}
        \label{fig:kvs-choreographic-flow}
    \end{subfigure}
    \caption{Comparison of the flow of control between the handwritten and choreographic KVS}
    \Description{Comparison of the flow of control between the handwritten and choreographic KVS}
    \label{fig:kvs-flow}
\end{figure}

While the protocol is simple, implementing it is error-prone. \Cref{fig:kvs-handwritten-flow} shows the implementation of the protocol \emph{without} using choreographic programming. The code defines three functions for each node. The highlight and arrows show the flow of data between the nodes. Because sends and receives are interleaved, it is difficult to track the flow of data and control.

\Cref{fig:kvs-choreographic-flow} shows the implementation of the same protocol as a choreography in ChoRus. The choreography communicates the request from \rs{Client} to \rs{Primary} using \rs{comm}. Then, it uses the \rs{enclave} operator to call the \rs{DoBackup} sub-choreography at \rs{Primary} and \rs{Backup}. The sub-choreography branches on the type of the request, and if the request is \rs{put}, it forwards the request to the backup node. After the sub-choreography returns, the primary node processes the request and returns the response to the client. The choreographic version is easier to understand because both data and control naturally flow from top to bottom.

While we could implement the KVS protocol as a choreography in HasChor, the naive implementation of conditionals in HasChor would present a problem. When we branch on the type of the request on the primary node, it broadcasts the type, and in HasChor, this broadcast would also go to the client, leaking an implementation detail.  By using enclaves, we can implement the protocol in a more efficient (and secure) manner.

\subsection{Case Study 2: Multiplayer Tic-Tac-Toe}
\label{subsec:eval-ttt}

An advantage of library-level choreographic programming is that it allows developers to reuse existing code. This is especially useful for implementing a distributed version of an existing local program. In this case study, we implement a distributed version of a tic-tac-toe game using ChoRus. We start with a local implementation of the game where two players play on the same computer. Then, we use ChoRus to port the local implementation to a distributed version, which lets the players play on different computers over the network, with minimal changes to the code. Finally, we compare the ChoRus implementation with a handwritten distributed version of the game.

\begin{figure}
    \includegraphics[width=\textwidth]{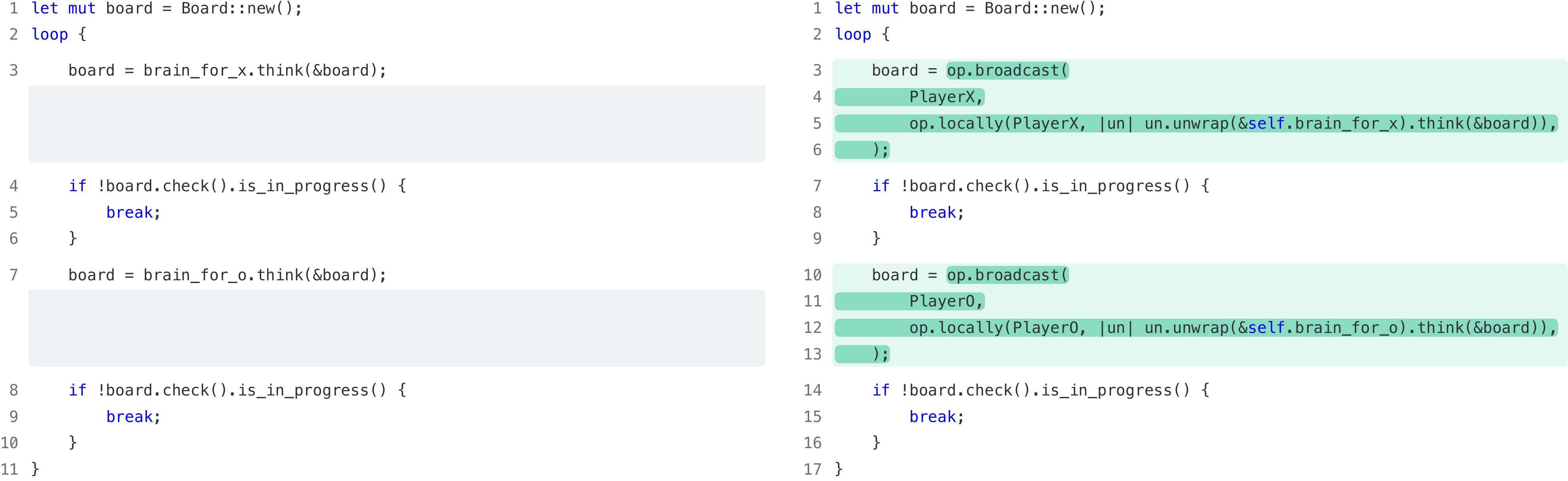}
    \caption{Diff between the local Rust and distributed ChoRus implementations of the tic-tac-toe game}
    \Description{Diff between the local Rust and distributed ChoRus implementations of the tic-tac-toe game}
    \label{fig:tic-tac-toe-diff}
\end{figure}

Let us start with the local implementation of the game. The left side of \Cref{fig:tic-tac-toe-diff} shows the structure of the main game loop written in Rust.
We omit the definitions of the structs and traits that capture the core logic of the game, such as \rs{board}, \rs{brain_for_x}, and \rs{brain_for_o}.
The game starts with an empty board. Then, the game enters a loop in which the two players take turns to make a move. After each move, we check the status of the board, and if the game is over, we break out of the loop. Finally, we print the result of the game.

Because ChoRus is a library, we can reuse the existing local Rust code to implement the distributed version of the game. The right side of \Cref{fig:tic-tac-toe-diff} shows the distributed implementation of the game written as a choreography in ChoRus. For brevity, we only show the \rs{run} method of the choreography. Just like its local counterpart, the choreography starts with an empty board. Then, the choreography enters a loop in which the two players make a local move and broadcast the new board. After each move, we check the status of the board, and if the game is over, we break out of the loop. Finally, we print the result of the game from the perspective of each player.

As highlighted in \Cref{fig:tic-tac-toe-diff}, changing the local implementation to the distributed implementation requires minimal changes to the code. All game logic and control flow are reused, and the only changes are the addition of the \rs{locally} and \rs{broadcast} operators to specify the location of data and computation. This is a significant advantage of library-level CP as opposed to a standalone CP language, because it allows developers to reuse existing code for local computation and focus on the distributed aspects of the program.

\subsection{Performance}
\label{subsec:eval-perf}

To employ CP in production, the performance overhead of using CP must be acceptable.  Performance is a particular concern for library-level CP, which involves carrying out EPP at runtime.  In this section, we measure the performance overhead of using ChoRus compared to traditional distributed programming in Rust. We focus on the overhead of running a choreography with EPP-as-DI. We conducted two experiments. First, we performed microbenchmarking to measure the overhead of EPP-as-DI in isolation. Second, we measured and compared the performance of the two versions of the key-value store from \Cref{subsec:eval-kvs}. All experiments in this section were performed on a MacBook Pro 2020 with an Apple M1 chip, 16 GB of RAM, and macOS Sonoma 14.0.

\subsubsection{Microbenchmarks}

With microbenchmarking, we measured the performance overhead of using two of the choreographic operators in ChoRus: \rs{locally} and \rs{comm}.

\begin{figure}
    \centering
    \begin{subfigure}[c]{.33\textwidth}
        \centering
        \includegraphics[width=\textwidth]{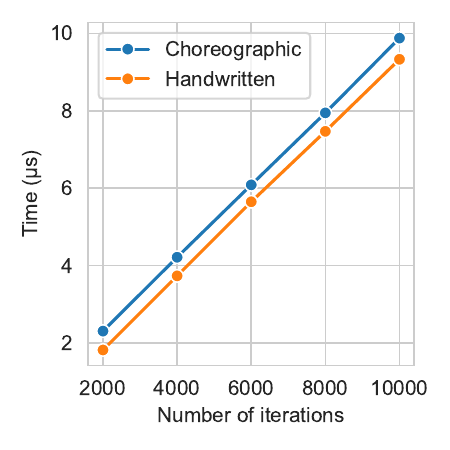}
        \caption{locally}
        \label{fig:microbenchmark-locally}
    \end{subfigure}%
    \begin{subfigure}[c]{.33\textwidth}
        \centering
        \includegraphics[width=\textwidth]{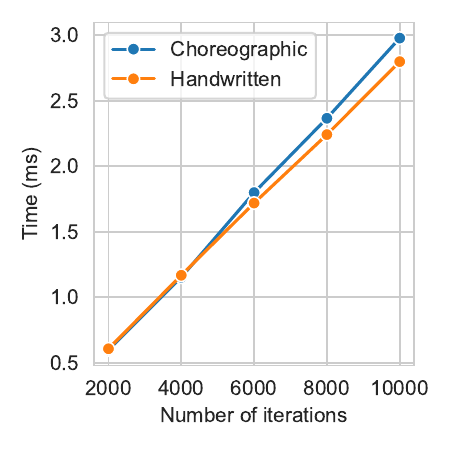}
        \caption{comm}
        \label{fig:microbenchmark-comm}
    \end{subfigure}%
    \begin{subfigure}[c]{.33\textwidth}
        \centering
        \includegraphics[width=\textwidth]{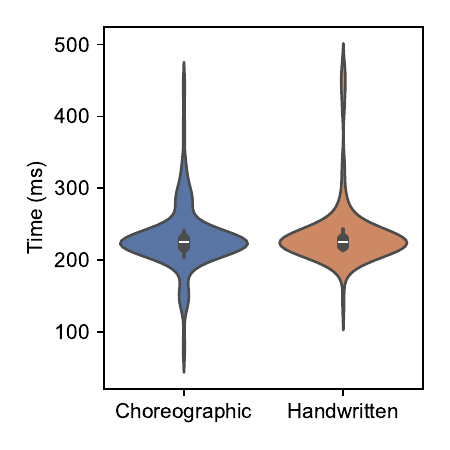}
        \caption{KVS Benchmark Results}
        \Description{KVS Benchmark Results}
        \label{fig:kvs-benchmark-results}
    \end{subfigure}
    \caption{Benchmark Results}
    \Description{Benchmark Results}
    \label{fig:benchmark-results}
\end{figure}

To measure the overhead of the \rs{locally} operator, we implemented a simple counter program as a handwritten Rust program and as a ChoRus choreography. The program initializes a counter and repeatedly increments it a given number of times. The ChoRus version is written as a choreography that runs only at one location and uses the \rs{locally} operator to perform initialization and increments. We use endpoint projection to execute the choreography. We measured the runtime of the two versions of the program with different numbers of iterations. \Cref{fig:microbenchmark-locally} shows the result of the microbenchmark. There is a small, constant overhead of using ChoRus. Because the overhead does not grow with the number of iterations, the overhead is likely due to the cost of endpoint projection, and there is no observable overhead of using the \rs{locally} operator.

We also measured the performance of the \rs{comm} operator. We implemented a simple protocol that moves data from one location to another as a handwritten Rust program and as a ChoRus choreography. In both the handwritten Rust and the ChoRus versions, to isolate the performance overhead of using EPP, we used ChoRus' \rs{LocalTransport} message transport layer to send data between the two locations.
\Cref{fig:microbenchmark-comm} shows the result. The message passing is dominating the running time in both versions, and the overhead of endpoint projection is not observable. The ChoRus version performed slightly worse for larger iterations, with a difference of <0.5 ms.

\subsubsection{Key-Value Store Benchmark}

We also benchmarked the two versions of the key-value store from \Cref{subsec:eval-kvs} to measure the system-level performance overhead of using ChoRus . We generated 100 random requests of 50\% \rs{get} and 50\% \rs{put} requests. We measured the runtime of the two versions of the program. We used \rs{HttpTransport} for communication between nodes in both versions. \Cref{fig:kvs-benchmark-results} shows a violin plot of 100 runs of the benchmark. The median runtime of the choreographic version was 225.09 ms, while the median runtime of the handwritten version was 224.93 ms. Even though the nodes are running on the same computer, the running time is dominated by the network latency, and we did not observe significant overhead of using ChoRus.

% !TEX root = ./main.tex

\section{Related Work}
\label{sec:related-work}

Choreographies were originally a specification mechanism for distributed systems~\citep{w3c-cdl-primer}. Researchers soon began to explore the notion of endpoint projection for choreographies~\citep{mendling-hafner-epp,qiu-foundation,carbone-cdl-epp-esop,carbone-cdl-epp,Lanese2008BridgingTG, McCarthy2008CryptographicPE}. The Chor language~\citep{deadlock-freedom-by-design, montesi-dissertation} pioneered the use of choreographies as executable programs by means of endpoint projection. Much of the subsequent literature on CP places emphasis on its formal foundations ~\citep{cruz-filipe-formalising-coq,hirsch2022pirouette, pohjola-kalas, cruz2022functional, graversen2023alice} rather than on practically usable implementations.
An exception to this rule is Choral~\cite{giallorenzo-choral}, arguably the most practical option among existing CP languages.  Choral has been used to implement widely used real-world protocols as choreographies~\citep{lugovic2023real}.
ChoRus and Choral share practicality and accessibility as goals, but make different design decisions in service of those goals.
As a Rust library, ChoRus enjoys the advantages of library-level CP discussed in \Cref{sec:intro}; as a standalone language with its own compiler, Choral has limited tooling and IDE support, although it prioritizes interoperability with Java, the target language for its EPP.  On the other hand, as a standalone CP language, Choral can statically generate local code for each endpoint to run, whereas ChoRus (and HasChor~\citep{HasChor}, the only other library-level CP implementation, which we discussed in \Cref{subsec:llcp}) must dynamically generate node-local programs at run time.

ChoRus and Choral differ in their treatment of message transport. Choral provides a hierarchy of \emph{channel types}, a very flexible abstraction that enables use of many types of user-defined channels within a single choreography.
The closest counterpart in ChoRus is the \rs{Transport} trait.  While users may implement their own \rs{Transport} types, they must choose a single implementation of \rs{Transport} for each \rs{Projector}.  This design decision is orthogonal to ChoRus's implementation of CP as a library, and a Choral-like channel abstraction would be feasible at the library level.  For our ChoRus case studies so far, though, the existing \rs{Transport} mechanism has sufficed.

ChoRus and Choral also handle propagation of knowledge of choice differently.  As we have seen, ChoRus supports choreographic enclaves~(\Cref{sec:enclaves}), which improve on the naive broadcast-based approach used in previous library-level CP implementations~\citep{HasChor}.  Choral, like other choreographic languages before it~\citep{deadlock-freedom-by-design, montesi-dissertation}, supports selection annotations via the \code{select} method.  Although enclaves are more fine-grained than naive broadcast, selection annotations give the programmer even more fine-grained control over how knowledge of choice is propagated.
Not all standalone CP languages use selection annotations; for instance, the language AIOC~\citep{dalla2014aiocj,aiocj} uses an endpoint projection approach that automatically broadcasts to the locations involved in both branches of a conditional expression --- a granularity similar to what enclaves offer, made possible by the ability to carry out static analysis on the branches.

ScalaLoci~\citep{weisenburger2018distributed} is a language for \emph{multitier} programming~\citep{weisenburger2020survey}, implemented as an embedded DSL in Scala.  Multitier programming can be seen as a close relative of CP~\citep{giallorenzo2021multiparty}, and ChoRus and ScalaLoci are both ``library-level'': a ScalaLoci program is a Scala program, just as a ChoRus program is a Rust program.
We observe that ScalaLoci --- and, perhaps, multitier programming generally --- seems to be a natural fit for asynchronous, reactive distributed applications, while ChoRus --- and perhaps CP generally --- seems to be a natural fit for synchronous, turn-based distributed applications, such as our tic-tac-toe case study or password authentication example.
We posit that the library level is a fruitful place to continue to explore the evident relationship between multitier and choreographic programming.

HasChor~\citep{HasChor} uses freer monads to embed CP in Haskell. Another popular technique to implement eDSLs in functional languages is \emph{tagless final} \citep{10.1007/978-3-540-76637-7_15}, where programmers use typeclasses to express abstract operations whose implementations are provided implicitly as typeclass instances.  Our EPP-as-DI technique is similar in spirit, but uses higher-order functions to explicitly pass choreographic operators as function arguments.

\section{Conclusion}

Library-level choreographic programming holds great promise.  In this paper, we aim to democratize library-level CP with techniques that make CP possible, efficient, and practical to implement in a wide variety of host languages. In particular, we presented \emph{endpoint projection as dependency injection (EPP-as-DI)}, a technique for library-level implementation of EPP in any host language that supports higher-order functions, and \emph{choreographic enclaves}, a language feature that improves on the efficiency of existing library-level CP.  We have implemented EPP-as-DI and choreographic enclaves in ChoRus, a library for CP in Rust. Our case studies and benchmarks demonstrate that programming wtih ChoRus compares favorably with traditional distributed programming in Rust. We hope that in the future, our proposed techniques will bring CP libraries to many host languages, and a subsequent expansion and diversification of CP in many communities.

\bibliographystyle{ACM-Reference-Format}
\bibliography{references}


\begin{thebibliography}{33}

%%% the Web address.

\ifx \showCODEN    \undefined \def \showCODEN     #1{\unskip}     \fi
\ifx \showDOI      \undefined \def \showDOI       #1{#1}\fi
\ifx \showISBNx    \undefined \def \showISBNx     #1{\unskip}     \fi
\ifx \showISBNxiii \undefined \def \showISBNxiii  #1{\unskip}     \fi
\ifx \showISSN     \undefined \def \showISSN      #1{\unskip}     \fi
\ifx \showLCCN     \undefined \def \showLCCN      #1{\unskip}     \fi
\ifx \shownote     \undefined \def \shownote      #1{#1}          \fi
\ifx \showarticletitle \undefined \def \showarticletitle #1{#1}   \fi
\ifx \showURL      \undefined \def \showURL       {\relax}        \fi
% invisible to TeX
\providecommand\bibfield[2]{#2}
\providecommand\bibinfo[2]{#2}
\providecommand\natexlab[1]{#1}
\providecommand\showeprint[2][]{arXiv:#2}

\bibitem[Basu and Bultan(2016)]%
        {basu-choreography-repair}
\bibfield{author}{\bibinfo{person}{Samik Basu} {and} \bibinfo{person}{Tevfik
  Bultan}.} \bibinfo{year}{2016}\natexlab{}.
\newblock \showarticletitle{Automated Choreography Repair}. In
  \bibinfo{booktitle}{\emph{Fundamental Approaches to Software Engineering}},
  \bibfield{editor}{\bibinfo{person}{Perdita Stevens} {and}
  \bibinfo{person}{Andrzej W{\k{a}}sowski}} (Eds.).
  \bibinfo{publisher}{Springer Berlin Heidelberg}, \bibinfo{address}{Berlin,
  Heidelberg}, \bibinfo{pages}{13--30}.
\newblock
\showISBNx{978-3-662-49665-7}


\bibitem[Carbone et~al\mbox{.}(2007)]%
        {carbone-cdl-epp-esop}
\bibfield{author}{\bibinfo{person}{Marco Carbone}, \bibinfo{person}{Kohei
  Honda}, {and} \bibinfo{person}{Nobuko Yoshida}.}
  \bibinfo{year}{2007}\natexlab{}.
\newblock \showarticletitle{Structured Communication-Centred Programming for
  Web Services}. In \bibinfo{booktitle}{\emph{Programming Languages and
  Systems}}, \bibfield{editor}{\bibinfo{person}{Rocco De~Nicola}} (Ed.).
  \bibinfo{publisher}{Springer Berlin Heidelberg}, \bibinfo{address}{Berlin,
  Heidelberg}, \bibinfo{pages}{2--17}.
\newblock
\showISBNx{978-3-540-71316-6}


\bibitem[Carbone et~al\mbox{.}(2012)]%
        {carbone-cdl-epp}
\bibfield{author}{\bibinfo{person}{Marco Carbone}, \bibinfo{person}{Kohei
  Honda}, {and} \bibinfo{person}{Nobuko Yoshida}.}
  \bibinfo{year}{2012}\natexlab{}.
\newblock \showarticletitle{Structured Communication-Centered Programming for
  Web Services}.
\newblock \bibinfo{journal}{\emph{ACM Trans. Program. Lang. Syst.}}
  \bibinfo{volume}{34}, \bibinfo{number}{2}, Article \bibinfo{articleno}{8}
  (\bibinfo{date}{June} \bibinfo{year}{2012}), \bibinfo{numpages}{78}~pages.
\newblock
\showISSN{0164-0925}
\urldef\tempurl%
\url{https://doi.org/10.1145/2220365.2220367}
\showDOI{\tempurl}


\bibitem[Carbone and Montesi(2013)]%
        {deadlock-freedom-by-design}
\bibfield{author}{\bibinfo{person}{Marco Carbone} {and}
  \bibinfo{person}{Fabrizio Montesi}.} \bibinfo{year}{2013}\natexlab{}.
\newblock \showarticletitle{Deadlock-Freedom-by-Design: Multiparty Asynchronous
  Global Programming}. In \bibinfo{booktitle}{\emph{Proceedings of the 40th
  Annual ACM SIGPLAN-SIGACT Symposium on Principles of Programming Languages}}
  (Rome, Italy) \emph{(\bibinfo{series}{POPL '13})}.
  \bibinfo{publisher}{Association for Computing Machinery},
  \bibinfo{address}{New York, NY, USA}, \bibinfo{pages}{263–274}.
\newblock
\showISBNx{9781450318327}
\urldef\tempurl%
\url{https://doi.org/10.1145/2429069.2429101}
\showDOI{\tempurl}


\bibitem[Carette et~al\mbox{.}(2007)]%
        {10.1007/978-3-540-76637-7_15}
\bibfield{author}{\bibinfo{person}{Jacques Carette}, \bibinfo{person}{Oleg
  Kiselyov}, {and} \bibinfo{person}{Chung-chieh Shan}.}
  \bibinfo{year}{2007}\natexlab{}.
\newblock \showarticletitle{Finally Tagless, Partially Evaluated}. In
  \bibinfo{booktitle}{\emph{Programming Languages and Systems}},
  \bibfield{editor}{\bibinfo{person}{Zhong Shao}} (Ed.).
  \bibinfo{publisher}{Springer Berlin Heidelberg}, \bibinfo{address}{Berlin,
  Heidelberg}, \bibinfo{pages}{222--238}.
\newblock
\showISBNx{978-3-540-76637-7}


\bibitem[Castagna et~al\mbox{.}(2011)]%
        {castagna-knowledge-of-choice}
\bibfield{author}{\bibinfo{person}{Giuseppe Castagna},
  \bibinfo{person}{Mariangiola Dezani-Ciancaglini}, {and} \bibinfo{person}{Luca
  Padovani}.} \bibinfo{year}{2011}\natexlab{}.
\newblock \showarticletitle{On Global Types and Multi-party Sessions}. In
  \bibinfo{booktitle}{\emph{Formal Techniques for Distributed Systems}},
  \bibfield{editor}{\bibinfo{person}{Roberto Bruni} {and}
  \bibinfo{person}{Juergen Dingel}} (Eds.). \bibinfo{publisher}{Springer Berlin
  Heidelberg}, \bibinfo{address}{Berlin, Heidelberg}, \bibinfo{pages}{1--28}.
\newblock
\showISBNx{978-3-642-21461-5}


\bibitem[Cruz-Filipe et~al\mbox{.}(2022)]%
        {cruz2022functional}
\bibfield{author}{\bibinfo{person}{Lu{\'\i}s Cruz-Filipe}, \bibinfo{person}{Eva
  Graversen}, \bibinfo{person}{Lovro Lugovi{\'c}}, \bibinfo{person}{Fabrizio
  Montesi}, {and} \bibinfo{person}{Marco Peressotti}.}
  \bibinfo{year}{2022}\natexlab{}.
\newblock \showarticletitle{Functional choreographic programming}. In
  \bibinfo{booktitle}{\emph{International Colloquium on Theoretical Aspects of
  Computing}}. Springer, \bibinfo{pages}{212--237}.
\newblock


\bibitem[Cruz-Filipe and Montesi(2020)]%
        {cruz-filipe-montesi-core}
\bibfield{author}{\bibinfo{person}{Luís Cruz-Filipe} {and}
  \bibinfo{person}{Fabrizio Montesi}.} \bibinfo{year}{2020}\natexlab{}.
\newblock \showarticletitle{A core model for choreographic programming}.
\newblock \bibinfo{journal}{\emph{Theoretical Computer Science}}
  \bibinfo{volume}{802} (\bibinfo{year}{2020}), \bibinfo{pages}{38--66}.
\newblock
\showISSN{0304-3975}
\urldef\tempurl%
\url{https://doi.org/10.1016/j.tcs.2019.07.005}
\showDOI{\tempurl}


\bibitem[Cruz-Filipe and Montesi(2023)]%
        {cruz-filipe-certified-repair}
\bibfield{author}{\bibinfo{person}{Lu{\'\i}s Cruz-Filipe} {and}
  \bibinfo{person}{Fabrizio Montesi}.} \bibinfo{year}{2023}\natexlab{}.
\newblock \showarticletitle{{Now It Compiles! Certified Automatic Repair of
  Uncompilable Protocols}}. In \bibinfo{booktitle}{\emph{14th International
  Conference on Interactive Theorem Proving (ITP 2023)}}
  \emph{(\bibinfo{series}{Leibniz International Proceedings in Informatics
  (LIPIcs)}, Vol.~\bibinfo{volume}{268})},
  \bibfield{editor}{\bibinfo{person}{Adam Naumowicz} {and}
  \bibinfo{person}{Ren\'{e} Thiemann}} (Eds.). \bibinfo{publisher}{Schloss
  Dagstuhl -- Leibniz-Zentrum f{\"u}r Informatik}, \bibinfo{address}{Dagstuhl,
  Germany}, \bibinfo{pages}{11:1--11:19}.
\newblock
\showISBNx{978-3-95977-284-6}
\showISSN{1868-8969}
\urldef\tempurl%
\url{https://doi.org/10.4230/LIPIcs.ITP.2023.11}
\showDOI{\tempurl}


\bibitem[Cruz-Filipe et~al\mbox{.}(2021)]%
        {cruz-filipe-formalising-coq}
\bibfield{author}{\bibinfo{person}{Lu{\'\i}s Cruz-Filipe},
  \bibinfo{person}{Fabrizio Montesi}, {and} \bibinfo{person}{Marco
  Peressotti}.} \bibinfo{year}{2021}\natexlab{}.
\newblock \showarticletitle{{Formalising a Turing-Complete Choreographic
  Language in Coq}}. In \bibinfo{booktitle}{\emph{12th International Conference
  on Interactive Theorem Proving (ITP 2021)}} \emph{(\bibinfo{series}{Leibniz
  International Proceedings in Informatics (LIPIcs)},
  Vol.~\bibinfo{volume}{193})}, \bibfield{editor}{\bibinfo{person}{Liron Cohen}
  {and} \bibinfo{person}{Cezary Kaliszyk}} (Eds.). \bibinfo{publisher}{Schloss
  Dagstuhl -- Leibniz-Zentrum f{\"u}r Informatik}, \bibinfo{address}{Dagstuhl,
  Germany}, \bibinfo{pages}{15:1--15:18}.
\newblock
\showISBNx{978-3-95977-188-7}
\showISSN{1868-8969}
\urldef\tempurl%
\url{https://doi.org/10.4230/LIPIcs.ITP.2021.15}
\showDOI{\tempurl}


\bibitem[{Dalla Preda} et~al\mbox{.}(2017)]%
        {aiocj}
\bibfield{author}{\bibinfo{person}{Mila {Dalla Preda}},
  \bibinfo{person}{Maurizio Gabbrielli}, \bibinfo{person}{Saverio Giallorenzo},
  \bibinfo{person}{Ivan Lanese}, {and} \bibinfo{person}{Jacopo Mauro}.}
  \bibinfo{year}{2017}\natexlab{}.
\newblock \showarticletitle{{Dynamic Choreographies: Theory And
  Implementation}}.
\newblock \bibinfo{journal}{\emph{{Logical Methods in Computer Science}}}
  \bibinfo{volume}{{Volume 13, Issue 2}} (\bibinfo{date}{April}
  \bibinfo{year}{2017}).
\newblock
\urldef\tempurl%
\url{https://doi.org/10.23638/LMCS-13(2:1)2017}
\showDOI{\tempurl}


\bibitem[{Dalla Preda} et~al\mbox{.}(2014)]%
        {dalla2014aiocj}
\bibfield{author}{\bibinfo{person}{Mila {Dalla Preda}},
  \bibinfo{person}{Saverio Giallorenzo}, \bibinfo{person}{Ivan Lanese},
  \bibinfo{person}{Jacopo Mauro}, {and} \bibinfo{person}{Maurizio Gabbrielli}.}
  \bibinfo{year}{2014}\natexlab{}.
\newblock \showarticletitle{AIOCJ: A choreographic framework for safe adaptive
  distributed applications}. In \bibinfo{booktitle}{\emph{Software Language
  Engineering: 7th International Conference, SLE 2014, V{\"a}ster{\aa}s,
  Sweden, September 15-16, 2014. Proceedings 7}}. Springer,
  \bibinfo{pages}{161--170}.
\newblock


\bibitem[Fowler(2004)]%
        {Fowler_2004}
\bibfield{author}{\bibinfo{person}{Martin Fowler}.}
  \bibinfo{year}{2004}\natexlab{}.
\newblock \bibinfo{title}{Inversion of control containers and the dependency
  injection pattern}.
\newblock
\newblock
\urldef\tempurl%
\url{https://martinfowler.com/articles/injection.html}
\showURL{%
\tempurl}


\bibitem[Giallorenzo et~al\mbox{.}(2020)]%
        {giallorenzo-choral}
\bibfield{author}{\bibinfo{person}{Saverio Giallorenzo},
  \bibinfo{person}{Fabrizio Montesi}, {and} \bibinfo{person}{Marco
  Peressotti}.} \bibinfo{year}{2020}\natexlab{}.
\newblock \bibinfo{title}{Object-Oriented Choreographic Programming}.
\newblock
\newblock
\urldef\tempurl%
\url{https://doi.org/10.48550/ARXIV.2005.09520}
\showDOI{\tempurl}


\bibitem[Giallorenzo et~al\mbox{.}(2021)]%
        {giallorenzo2021multiparty}
\bibfield{author}{\bibinfo{person}{Saverio Giallorenzo},
  \bibinfo{person}{Fabrizio Montesi}, \bibinfo{person}{Marco Peressotti},
  \bibinfo{person}{David Richter}, \bibinfo{person}{Guido Salvaneschi}, {and}
  \bibinfo{person}{Pascal Weisenburger}.} \bibinfo{year}{2021}\natexlab{}.
\newblock \showarticletitle{Multiparty languages: The choreographic and
  multitier cases}. In \bibinfo{booktitle}{\emph{ECOOP 2021-European Conference
  on Object-Oriented Programming}}.
\newblock


\bibitem[Graversen et~al\mbox{.}(2023)]%
        {graversen2023alice}
\bibfield{author}{\bibinfo{person}{Eva Graversen}, \bibinfo{person}{Andrew~K
  Hirsch}, {and} \bibinfo{person}{Fabrizio Montesi}.}
  \bibinfo{year}{2023}\natexlab{}.
\newblock \showarticletitle{Alice or Bob?: Process Polymorphism in
  Choreographies}.
\newblock \bibinfo{journal}{\emph{arXiv preprint arXiv:2303.04678}}
  (\bibinfo{year}{2023}).
\newblock


\bibitem[Hirsch and Garg(2022)]%
        {hirsch2022pirouette}
\bibfield{author}{\bibinfo{person}{Andrew~K Hirsch} {and}
  \bibinfo{person}{Deepak Garg}.} \bibinfo{year}{2022}\natexlab{}.
\newblock \showarticletitle{Pirouette: higher-order typed functional
  choreographies}.
\newblock \bibinfo{journal}{\emph{Proceedings of the ACM on Programming
  Languages}} \bibinfo{volume}{6}, \bibinfo{number}{POPL}
  (\bibinfo{year}{2022}), \bibinfo{pages}{1--27}.
\newblock


\bibitem[Honda et~al\mbox{.}(2008)]%
        {honda-mpsts}
\bibfield{author}{\bibinfo{person}{Kohei Honda}, \bibinfo{person}{Nobuko
  Yoshida}, {and} \bibinfo{person}{Marco Carbone}.}
  \bibinfo{year}{2008}\natexlab{}.
\newblock \showarticletitle{Multiparty Asynchronous Session Types}. In
  \bibinfo{booktitle}{\emph{Proceedings of the 35th Annual ACM SIGPLAN-SIGACT
  Symposium on Principles of Programming Languages}} (San Francisco,
  California, USA) \emph{(\bibinfo{series}{POPL '08})}.
  \bibinfo{publisher}{Association for Computing Machinery},
  \bibinfo{address}{New York, NY, USA}, \bibinfo{pages}{273–284}.
\newblock
\showISBNx{9781595936899}
\urldef\tempurl%
\url{https://doi.org/10.1145/1328438.1328472}
\showDOI{\tempurl}


\bibitem[Hudak(1996)]%
        {hudak-edsls}
\bibfield{author}{\bibinfo{person}{Paul Hudak}.}
  \bibinfo{year}{1996}\natexlab{}.
\newblock \showarticletitle{Building Domain-Specific Embedded Languages}.
\newblock \bibinfo{journal}{\emph{ACM Comput. Surv.}} \bibinfo{volume}{28},
  \bibinfo{number}{4es} (\bibinfo{date}{dec} \bibinfo{year}{1996}),
  \bibinfo{pages}{196–es}.
\newblock
\showISSN{0360-0300}
\urldef\tempurl%
\url{https://doi.org/10.1145/242224.242477}
\showDOI{\tempurl}


\bibitem[Kiselyov and Ishii(2015)]%
        {kiselyov-more-ext-effs}
\bibfield{author}{\bibinfo{person}{Oleg Kiselyov} {and} \bibinfo{person}{Hiromi
  Ishii}.} \bibinfo{year}{2015}\natexlab{}.
\newblock \showarticletitle{Freer Monads, More Extensible Effects}. In
  \bibinfo{booktitle}{\emph{Proceedings of the 2015 ACM SIGPLAN Symposium on
  Haskell}} (Vancouver, BC, Canada) \emph{(\bibinfo{series}{Haskell '15})}.
  \bibinfo{publisher}{Association for Computing Machinery},
  \bibinfo{address}{New York, NY, USA}, \bibinfo{pages}{94–105}.
\newblock
\showISBNx{9781450338080}
\urldef\tempurl%
\url{https://doi.org/10.1145/2804302.2804319}
\showDOI{\tempurl}


\bibitem[Lanese et~al\mbox{.}(2008)]%
        {Lanese2008BridgingTG}
\bibfield{author}{\bibinfo{person}{Ivan Lanese}, \bibinfo{person}{Claudio
  Guidi}, \bibinfo{person}{Fabrizio Montesi}, {and} \bibinfo{person}{Gianluigi
  Zavattaro}.} \bibinfo{year}{2008}\natexlab{}.
\newblock \showarticletitle{Bridging the Gap between Interaction- and
  Process-Oriented Choreographies}.
\newblock \bibinfo{journal}{\emph{2008 Sixth IEEE International Conference on
  Software Engineering and Formal Methods}} (\bibinfo{year}{2008}),
  \bibinfo{pages}{323--332}.
\newblock
\urldef\tempurl%
\url{https://api.semanticscholar.org/CorpusID:11388743}
\showURL{%
\tempurl}


\bibitem[Lanese et~al\mbox{.}(2013)]%
        {lanese-amending-choreographies}
\bibfield{author}{\bibinfo{person}{Ivan Lanese}, \bibinfo{person}{Fabrizio
  Montesi}, {and} \bibinfo{person}{Gianluigi Zavattaro}.}
  \bibinfo{year}{2013}\natexlab{}.
\newblock \showarticletitle{Amending Choreographies}. In
  \bibinfo{booktitle}{\emph{Proceedings 9th International Workshop on Automated
  Specification and Verification of Web Systems, {WWV} 2013, Florence, Italy,
  6th June 2013}} \emph{(\bibinfo{series}{{EPTCS}},
  Vol.~\bibinfo{volume}{123})},
  \bibfield{editor}{\bibinfo{person}{Ant{\'{o}}nio Ravara} {and}
  \bibinfo{person}{Josep Silva}} (Eds.). \bibinfo{pages}{34--48}.
\newblock
\urldef\tempurl%
\url{https://doi.org/10.4204/EPTCS.123.5}
\showDOI{\tempurl}


\bibitem[Lugovi{\'c} and Montesi(2023)]%
        {lugovic2023real}
\bibfield{author}{\bibinfo{person}{Lovro Lugovi{\'c}} {and}
  \bibinfo{person}{Fabrizio Montesi}.} \bibinfo{year}{2023}\natexlab{}.
\newblock \showarticletitle{Real-World Choreographic Programming: An Experience
  Report}.
\newblock \bibinfo{journal}{\emph{arXiv preprint arXiv:2303.03983}}
  (\bibinfo{year}{2023}).
\newblock


\bibitem[McCarthy and Krishnamurthi(2008)]%
        {McCarthy2008CryptographicPE}
\bibfield{author}{\bibinfo{person}{Jay~A. McCarthy} {and}
  \bibinfo{person}{Shriram Krishnamurthi}.} \bibinfo{year}{2008}\natexlab{}.
\newblock \showarticletitle{Cryptographic Protocol Explication and End-Point
  Projection}. In \bibinfo{booktitle}{\emph{European Symposium on Research in
  Computer Security}}.
\newblock
\urldef\tempurl%
\url{https://api.semanticscholar.org/CorpusID:15429446}
\showURL{%
\tempurl}


\bibitem[Mendling and Hafner(2005)]%
        {mendling-hafner-epp}
\bibfield{author}{\bibinfo{person}{Jan Mendling} {and} \bibinfo{person}{Michael
  Hafner}.} \bibinfo{year}{2005}\natexlab{}.
\newblock \showarticletitle{From Inter-Organizational Workflows to Process
  Execution: Generating BPEL from WS-CDL}. In
  \bibinfo{booktitle}{\emph{Proceedings of the 2005 OTM Confederated
  International Conference on On the Move to Meaningful Internet Systems}}
  (Agia Napa, Cyprus) \emph{(\bibinfo{series}{OTM'05})}.
  \bibinfo{publisher}{Springer-Verlag}, \bibinfo{address}{Berlin, Heidelberg},
  \bibinfo{pages}{506–515}.
\newblock
\showISBNx{3540297391}
\urldef\tempurl%
\url{https://doi.org/10.1007/11575863_70}
\showDOI{\tempurl}


\bibitem[Montesi(2013)]%
        {montesi-dissertation}
\bibfield{author}{\bibinfo{person}{Fabrizio Montesi}.}
  \bibinfo{year}{2013}\natexlab{}.
\newblock \emph{\bibinfo{title}{Choreographic {P}rogramming}}.
\newblock Ph.{D}. Thesis. \bibinfo{school}{IT University of Copenhagen}.
\newblock
\newblock
\shownote{\url{https://www.fabriziomontesi.com/files/choreographic-programming.pdf}}.


\bibitem[Montesi(2023)]%
        {montesi-book}
\bibfield{author}{\bibinfo{person}{Fabrizio Montesi}.}
  \bibinfo{year}{2023}\natexlab{}.
\newblock \bibinfo{booktitle}{\emph{Introduction to Choreographies}}.
\newblock \bibinfo{publisher}{Cambridge University Press}.
\newblock


\bibitem[Pohjola et~al\mbox{.}(2022)]%
        {pohjola-kalas}
\bibfield{author}{\bibinfo{person}{Johannes~\r{A}man Pohjola},
  \bibinfo{person}{Alejandro G\'{o}mez-Londo\~{n}o}, \bibinfo{person}{James
  Shaker}, {and} \bibinfo{person}{Michael Norrish}.}
  \bibinfo{year}{2022}\natexlab{}.
\newblock \showarticletitle{{Kalas: A Verified, End-To-End Compiler for a
  Choreographic Language}}. In \bibinfo{booktitle}{\emph{13th International
  Conference on Interactive Theorem Proving (ITP 2022)}}
  \emph{(\bibinfo{series}{Leibniz International Proceedings in Informatics
  (LIPIcs)}, Vol.~\bibinfo{volume}{237})},
  \bibfield{editor}{\bibinfo{person}{June Andronick} {and}
  \bibinfo{person}{Leonardo de~Moura}} (Eds.). \bibinfo{publisher}{Schloss
  Dagstuhl -- Leibniz-Zentrum f{\"u}r Informatik}, \bibinfo{address}{Dagstuhl,
  Germany}, \bibinfo{pages}{27:1--27:18}.
\newblock
\showISBNx{978-3-95977-252-5}
\showISSN{1868-8969}
\urldef\tempurl%
\url{https://doi.org/10.4230/LIPIcs.ITP.2022.27}
\showDOI{\tempurl}


\bibitem[Qiu et~al\mbox{.}(2007)]%
        {qiu-foundation}
\bibfield{author}{\bibinfo{person}{Zongyan Qiu}, \bibinfo{person}{Xiangpeng
  Zhao}, \bibinfo{person}{Chao Cai}, {and} \bibinfo{person}{Hongli Yang}.}
  \bibinfo{year}{2007}\natexlab{}.
\newblock \showarticletitle{Towards the Theoretical Foundation of
  Choreography}. In \bibinfo{booktitle}{\emph{Proceedings of the 16th
  International Conference on World Wide Web}} (Banff, Alberta, Canada)
  \emph{(\bibinfo{series}{WWW '07})}. \bibinfo{publisher}{Association for
  Computing Machinery}, \bibinfo{address}{New York, NY, USA},
  \bibinfo{pages}{973–982}.
\newblock
\showISBNx{9781595936547}
\urldef\tempurl%
\url{https://doi.org/10.1145/1242572.1242704}
\showDOI{\tempurl}


\bibitem[Shen et~al\mbox{.}(2023)]%
        {HasChor}
\bibfield{author}{\bibinfo{person}{Gan Shen}, \bibinfo{person}{Shun Kashiwa},
  {and} \bibinfo{person}{Lindsey Kuper}.} \bibinfo{year}{2023}\natexlab{}.
\newblock \showarticletitle{{HasChor: Functional Choreographic Programming for
  All}}.
\newblock \bibinfo{journal}{\emph{Proc. ACM Program. Lang.}}
  \bibinfo{volume}{7}, \bibinfo{number}{ICFP} (\bibinfo{date}{Aug.}
  \bibinfo{year}{2023}).
\newblock
\urldef\tempurl%
\url{https://doi.org/10.1145/3607849}
\showDOI{\tempurl}


\bibitem[Weisenburger et~al\mbox{.}(2018)]%
        {weisenburger2018distributed}
\bibfield{author}{\bibinfo{person}{Pascal Weisenburger}, \bibinfo{person}{Mirko
  K{\"o}hler}, {and} \bibinfo{person}{Guido Salvaneschi}.}
  \bibinfo{year}{2018}\natexlab{}.
\newblock \showarticletitle{Distributed system development with ScalaLoci}.
\newblock \bibinfo{journal}{\emph{Proceedings of the ACM on Programming
  Languages}} \bibinfo{volume}{2}, \bibinfo{number}{OOPSLA}
  (\bibinfo{year}{2018}), \bibinfo{pages}{1--30}.
\newblock


\bibitem[Weisenburger et~al\mbox{.}(2020)]%
        {weisenburger2020survey}
\bibfield{author}{\bibinfo{person}{Pascal Weisenburger},
  \bibinfo{person}{Johannes Wirth}, {and} \bibinfo{person}{Guido Salvaneschi}.}
  \bibinfo{year}{2020}\natexlab{}.
\newblock \showarticletitle{A survey of multitier programming}.
\newblock \bibinfo{journal}{\emph{ACM Computing Surveys (CSUR)}}
  \bibinfo{volume}{53}, \bibinfo{number}{4} (\bibinfo{year}{2020}),
  \bibinfo{pages}{1--35}.
\newblock


\bibitem[{World Wide Web Consortium}(2006)]%
        {w3c-cdl-primer}
\bibfield{author}{\bibinfo{person}{The {World Wide Web Consortium}}.}
  \bibinfo{year}{2006}\natexlab{}.
\newblock \bibinfo{title}{Web Services Choreography Description Language:
  Primer}.
\newblock
\newblock
\urldef\tempurl%
\url{https://www.w3.org/TR/ws-cdl-10-primer/}
\showURL{%
\tempurl}


\end{thebibliography}

\end{document}